\documentclass[10pt,twoside]{article}
\usepackage{artmods,macros,subeqn}
\usepackage{makeidx}
\makeindex

\begin{document}

  \let\Sec=\S

   \def\noproof{\hspace{1em}\blackslug}
  \def\mm#1{\hbox{$#1$}}                 
  \def\mtx#1#2{\renewcommand{\arraystretch}{1.2}%
      \left(\! \begin{array}{#1}#2\end{array}\! \right)}


\def\twonorm#1{\norm{#1}_2}
\def\onenorm#1{\norm{#1}_1}
\def\infnorm#1{\norm{#1}_{\infty}}
\def\disp{\displaystyle}
\def\m{\phantom-}
\def\EM#1{{\em#1\/}}
\def\ur{\mbox{\bf u}}
\def\Real{I\!\!R}
\def\Re{I\!\!R}
\def\FR{I\!\!F}
\def\Complex{I\mskip -11.3mu C}
\def\sp{\hspace{10pt}}
\def\subject{\hbox{\rm subject to}}
\def\minimize#1{{\displaystyle\minim_{#1}}} 
\def\minimizetwoarg#1#2#3#4{{\displaystyle\minim_{#1 \in \Re^{#2},%
                                                  #3 \in \Re^{#4}}}}
\def\maximize#1{{\displaystyle\maxim_{#1}}} 
\def\eval#1#2{\left.#2\right|_{\alpha=#1}}

\def\tmat#1#2{(\, #1 \ \ #2 \,)}
\def\tmatt#1#2#3{(\, #1 \ \  #2 \ \  #3\,)}
\def\tmattt#1#2#3#4{(\, #1 \ \  #2 \ \  #3 \ \  #4\,)}

\def\mat#1#2{(\; #1 \quad #2 \;)}
\def\matt#1#2#3{(\; #1 \quad #2 \quad #3\;)}
\def\mattt#1#2#3#4{(\; #1 \quad #2 \quad #3 \quad #4\;)}
\def\blackslug{\hbox{\hskip 1pt \vrule width 4pt height 6pt depth 1.5pt
  \hskip 1pt}}
\def\boxit#1{\vbox{\hrule\hbox{\vrule\hskip 3pt
  \vbox{\vskip 3pt #1 \vskip 3pt}\hskip 3pt\vrule}\hrule}}
\def\cross{\scriptscriptstyle\times}
\def\maxim{\mathop{\hbox{\rm maximize}}}
\def\minim{\mathop{\hbox{\rm minimize}}}
\def\cond{\hbox{\rm cond}}
\def\det{\mathop{\hbox{\rm det}}}
\def\diag{\mathop{\hbox{\rm diag}}}
\def\dim{\mathop{\hbox{\rm dim}}}  
\def\exp{\mathop{\hbox{\rm exp}}}  
\def\In{\mathop{\hbox{\it In}\,}}
\def\boundary{\mathop{\hbox{\rm bnd}}}
\def\interior{\mathop{\hbox{\rm int}}}
\def\Null{\mathop{\hbox{\rm null}}}
\def\rank{\mathop{\hbox{\rm rank}}}
\def\op{\mathop{\hbox{\rm op}}}
\def\Range{\mathop{\hbox{\rm range}}}
\def\range{\mathop{\hbox{\rm range}}}
\def\sign{\mathop{\hbox{\rm sign}}}
\def\sgn{\mathop{\hbox{\rm sgn}}}
\def\Span{\mbox{\rm span}}
\def\trace{\mathop{\hbox{\rm trace}}}
\def\drop{^{\null}}
\def\etal{{\em et al.\ }}  
\def\float{fl}
\def\grad{\nabla}
\def\Hess{\nabla^2}
\def\half  {{\textstyle{1\over 2}}}
\def\third {{\textstyle{1\over 3}}}
\def\fourth{{\textstyle{1\over 4}}}
\def\sixth{{\textstyle{1\over 6}}}
\def\inv{^{-1}}
\def\invsq{^{-2}}
\def\limk{\lim_{k\to\infty}}
\def\mystrut{\vrule height10.5pt depth5.5pt width0pt}
\def\mod#1{|#1|}
\def\modd#1{\biggl|#1\biggr|}
\def\Mod#1{\left|#1\right|}
\def\normm#1{\biggl\|#1\biggr\|}
\def\norm#1{\|#1\|}
\def\spose#1{\hbox to 0pt{#1\hss}}
\def\sub#1{^{\null}_{#1}}
\def\text #1{\hbox{\quad#1\quad}}
\def\textt#1{\hbox{\qquad#1\qquad}}
\def\T{^T\!}
\def\ntext#1{\noalign{\vskip 2pt\hbox{#1}\vskip 2pt}}
\def\enddef{{$\null$}}
\def\xint{{x_{\rm int}}}
\def\vhat{{\hat v}}


\def\pthinsp{\mskip  2   mu}    
\def\pmedsp {\mskip  2.75mu}    
\def\pthiksp{\mskip  3.5 mu}    
\def\nthinsp{\mskip -2   mu}
\def\nmedsp {\mskip -2.75mu}
\def\nthiksp{\mskip -3.5 mu}


\def\Asubk{A\sub{\nthinsp k}}
\def\Bsubk{B\sub{\nthinsp k}}
\def\Fsubk{F\sub{\nmedsp k}}
\def\Gsubk{G\sub{\nthinsp k}}
\def\Hsubk{H\sub{\nthinsp k}}
\def\HsubB{H\sub{\nthinsp \scriptscriptstyle{B}}}
\def\HsubS{H\sub{\nthinsp \scriptscriptstyle{S}}}
\def\HsubBS{H\sub{\nthinsp \scriptscriptstyle{BS}}}
\def\Hz{H\sub{\nthinsp \scriptscriptstyle{Z}}}
\def\Jsubk{J\sub{\nmedsp k}}
\def\Psubk{P\sub{\nthiksp k}}
\def\Qsubk{Q\sub{\nthinsp k}}
\def\Vsubk{V\sub{\nmedsp k}}
\def\Vsubone{V\sub{\nmedsp 1}}
\def\Vsubtwo{V\sub{\nmedsp 2}}
\def\Ysubk{Y\sub{\nmedsp k}}
\def\Wsubk{W\sub{\nthiksp k}}
\def\Zsubk{Z\sub{\nthinsp k}}
\def\Zsubi{Z\sub{\nthinsp i}}


\def\subfr{_{\scriptscriptstyle{\it FR}}}
\def\subfx{_{\scriptscriptstyle{\it FX}}}
\def\fr{_{\scriptscriptstyle{\it FR}}}
\def\fx{_{\scriptscriptstyle{\it FX}}}
\def\submax{_{\max}}
\def\submin{_{\min}}
\def\subminus{_{\scriptscriptstyle -}}
\def\subplus {_{\scriptscriptstyle +}}

\def\A{_{\scriptscriptstyle A}}
\def\B{_{\scriptscriptstyle B}}
\def\BS{_{\scriptscriptstyle BS}}
\def\C{_{\scriptscriptstyle C}}
\def\D{_{\scriptscriptstyle D}}
\def\E{_{\scriptscriptstyle E}}
\def\F{_{\scriptscriptstyle F}}
\def\G{_{\scriptscriptstyle G}}
\def\H{_{\scriptscriptstyle H}}
\def\I{_{\scriptscriptstyle I}}
\def\w{_w}
\def\k{_k}
\def\kp#1{_{k+#1}}
\def\km#1{_{k-#1}}
\def\j{_j}
\def\i{_i}
\def\jp#1{_{j+#1}}
\def\jm#1{_{j-#1}}
\def\K{_{\scriptscriptstyle K}}
\def\L{_{\scriptscriptstyle L}}
\def\M{_{\scriptscriptstyle M}}
\def\N{_{\scriptscriptstyle N}}
\def\subb{\B}
\def\subc{\C}
\def\subm{\M}
\def\subf{\F}
\def\subn{\N}
\def\subh{\H}
\def\subk{\K}
\def\subt{_{\scriptscriptstyle T}}
\def\subr{\R}
\def\subz{\Z}
\def\J{_{\scriptscriptstyle J}}
\def\O{_{\scriptscriptstyle O}}
\def\Q{_{\scriptscriptstyle Q}}
\def\P{_{\scriptscriptstyle P}}
\def\R{_{\scriptscriptstyle R}}
\def\S{_{\scriptscriptstyle S}}
\def\U{_{\scriptscriptstyle U}}
\def\V{_{\scriptscriptstyle V}}
\def\W{_{\scriptscriptstyle W}}
\def\Y{_{\scriptscriptstyle Y}}
\def\Z{_{\scriptscriptstyle Z}}


\def\superstar{^{\raise 0.5pt\hbox{$\nthinsp *$}}}
\def\SUPERSTAR{^{\raise 0.5pt\hbox{$*$}}}
\def\shrink{\mskip -7mu}  

\def\alphastar{\alpha\superstar}
\def\lamstar  {\lambda\superstar}
\def\lamstarT {\lambda^{\raise 0.5pt\hbox{$\nthinsp *$}T}}
\def\lambdastar{\lamstar}
\def\nustar{\nu\superstar}
\def\mustar{\mu\superstar}
\def\pistar{\pi\superstar}

\def\cstar{c\superstar}
\def\dstar{d\SUPERSTAR}
\def\fstar{f\SUPERSTAR}
\def\gstar{g\superstar}
\def\hstar{h\superstar}
\def\mstar{m\superstar}
\def\pstar{p\superstar}
\def\sstar{s\superstar}
\def\ustar{u\superstar}
\def\vstar{v\superstar}
\def\wstar{w\superstar}
\def\xstar{x\superstar}
\def\ystar{y\superstar}
\def\zstar{z\superstar}

\def\Astar{A\SUPERSTAR}
\def\Bstar{B\SUPERSTAR}
\def\Cstar{C\SUPERSTAR}
\def\Fstar{F\SUPERSTAR}
\def\Gstar{G\SUPERSTAR}
\def\Hstar{H\SUPERSTAR}
\def\Jstar{J\SUPERSTAR}
\def\Kstar{K\SUPERSTAR}
\def\Mstar{M\SUPERSTAR}
\def\Ustar{U\SUPERSTAR}
\def\Vstar{V\SUPERSTAR}
\def\Wstar{W\SUPERSTAR}
\def\Xstar{X\SUPERSTAR}
\def\Zstar{Z\SUPERSTAR}


\def\alphabar{\bar \alpha}
\def\alphahat{\skew3\hat \alpha}
\def\alphatilde{\skew3\tilde \alpha}
\def\betabar{\skew{2.8}\bar\beta}
\def\betahat{\skew{2.8}\hat\beta}
\def\betatilde{\skew{2.8}\tilde\beta}
\def\delbar{\bar\delta}
\def\deltilde{\skew5\tilde \delta}
\def\deltabar{\delbar}
\def\deltatilde{\deltilde}
\def\lambar{\bar\lambda}
\def\etabar{\bar\eta}
\def\gammabar{\bar\gamma}
\def\lamhat{\skew{2.8}\hat \lambda}
\def\lambdabar{\lambar}
\def\lambdahat{\lamhat}
\def\mubar{\skew3\bar \mu}
\def\muhat{\skew3\hat \mu}
\def\mutilde{\skew3\tilde\mu}
\def\nubar{\skew3\bar\nu}
\def\nuhat{\skew3\hat\nu}
\def\nutilde{\skew3\tilde\nu}
\def\omegabar{\bar\omega}
\def\pibar{\bar\pi}
\def\pihat{\skew1\widehat \pi}
\def\sigmabar{\bar\sigma}
\def\rhobar{\bar\rho}
\def\rhohat{\widehat\rho}
\def\taubar{\bar\tau}
\def\tautilde{\tilde\tau}
\def\tauhat{\hat\tau}
\def\thetabar{\bar\theta}
\def\xibar{\skew4\bar\xi}

\def\Deltait{{\mit \Delta}}
\def\Gammait{{\mit \Gamma}}
\def\Lambdait{{\mit \Lambda}}
\def\Sigmait{{\mit \Sigma}}
\def\Lambarit{\skew5\bar{\mit \Lambda}}
\def\Omegait{{\mit \Omega}}
\def\Omegaitbar{\skew5\bar{\mit \Omega}}
\def\Thetait{{\mit \Theta}}
\def\Piitbar{\skew5\bar{\mit \Pi}}
\def\Piit{{\mit \Pi}}
\def\Phiit{{\mit \Phi}}
\def\Ascr{{\cal A}}
\def\Bscr{{\cal B}}
\def\Fscr{{\cal F}}
\def\Dscr{{\cal D}}
\def\Iscr{{\cal I}}
\def\Jscr{{\cal J}}
\def\Lscr{{\cal L}}
\def\Mscr{{\cal M}}
\def\lscr{\ell}
\def\lscrbar{\ellbar}
\def\Oscr{{\cal O}}
\def\Pscr{{\cal P}}
\def\Qscr{{\cal Q}}
\def\Sscr{{\cal S}}
\def\Uscr{{\cal U}}
\def\Vscr{{\cal V}}
\def\Wscr{{\cal W}}
\def\Mscr{{\cal M}}
\def\Nscr{{\cal N}}
\def\Rscr{{\cal R}}

\def\abar{\skew3\bar a}
\def\ahat{\skew2\widehat a}
\def\atilde{\skew2\widetilde a}
\def\Abar{\skew7\bar A}
\def\Ahat{\widehat A}
\def\Atilde{\widetilde A}
\def\bbar{\skew3\bar b}
\def\bhat{\skew2\widehat b}
\def\btilde{\skew2\widetilde b}
\def\Bbar{\bar B}
\def\Bhat{\widehat B}
\def\cbar{\skew5\bar c}
\def\chat{\skew3\widehat c}
\def\ctilde{\widetilde c}
\def\Cbar{\bar C}
\def\Chat{\widehat C}
\def\Ctilde{\widetilde C}
\def\dbar{\bar d}
\def\dhat{\widehat d}
\def\dtilde{\widetilde d}
\def\Dbar{\bar D}
\def\Dhat{\widehat D}
\def\Dtilde{\widetilde D}
\def\ehat{\skew3\widehat e}
\def\ebar{\bar e}
\def\Ebar{\bar E}
\def\Ehat{\widehat E}
\def\fbar{\bar f}
\def\fhat{\widehat f}
\def\ftilde{\widetilde f}
\def\Fbar{\bar F}
\def\Fhat{\widehat F}
\def\gbar{\skew{4.3}\bar g}
\def\ghat{\skew{4.3}\widehat g}
\def\gtilde{\skew{4.5}\widetilde g}
\def\Gbar{\bar G}
\def\Ghat{\widehat G}
\def\hbar{\skew{4.2}\bar h}
\def\hhat{\skew2\widehat h}
\def\htilde{\skew3\widetilde h}
\def\Hbar{\skew5\bar H}
\def\Hhat{\widehat H}
\def\Htilde{\widetilde H}
\def\Ibar{\skew5\bar I}
\def\Itilde{\widetilde I}
\def\Jbar{\skew6\bar J}
\def\Jhat{\widehat J}
\def\Jtilde{\widetilde J}
\def\kbar{\skew{4.4}\bar k}
\def\Khat{\widehat K}
\def\Kbar{\skew{4.4}\bar K}
\def\Ktilde{\widetilde K}
\def\ellbar{\bar \ell}
\def\lhat{\skew2\widehat l}
\def\lbar{\skew2\bar l}
\def\Lbar{\skew{4.3}\bar L}
\def\Lhat{\widehat L}
\def\Ltilde{\widetilde L}
\def\mbar{\skew2\bar m}
\def\mhat{\widehat m}
\def\Mbar{\skew{4.4}\bar M}
\def\Mhat{\widehat M}
\def\Mtilde{\widetilde M}
\def\Nbar{\skew{4.4}\bar N}
\def\Ntilde{\widetilde N}
\def\nbar{\skew2\bar n}
\def\pbar{\skew2\bar p}
\def\phat{\skew2\widehat p}
\def\ptilde{\skew2\widetilde p}
\def\Pbar{\skew5\bar P}
\def\Phat{\widehat P}
\def\Ptilde{\skew5\widetilde P}
\def\qbar{\bar q}
\def\qhat{\skew2\widehat q}
\def\qtilde{\widetilde q}
\def\Qbar{\bar Q}
\def\Qhat{\widehat Q}
\def\Qtilde{\widetilde Q}
\def\rbar{\skew3\bar r}
\def\rhat{\skew3\widehat r}
\def\rtilde{\skew3\widetilde r}
\def\Rbar{\skew5\bar R}
\def\Rhat{\widehat R}
\def\Rtilde{\widetilde R}
\def\sbar{\bar s}
\def\shat{\widehat s}
\def\Stilde{\widetilde S}
\def\stilde{\widetilde s}
\def\Shat{\widehat S}
\def\Sbar{\skew2\bar S}
\def\tbar{\bar t}
\def\ttilde{\widetilde t}
\def\that{\widehat t}
\def\Tbar{\bar T}
\def\That{\widehat T}
\def\Ttilde{\widetilde T}
\def\ubar{\skew3\bar u}
\def\uhat{\skew3\widehat u}
\def\utilde{\skew3\widetilde u}
\def\Ubar{\skew2\bar U}
\def\Uhat{\widehat U}
\def\Utilde{\widetilde U}
\def\vbar{\skew3\bar v}
\def\vhat{\skew3\widehat v}
\def\vtilde{\skew3\widetilde v}
\def\Vbar{\skew2\bar V}
\def\Vhat{\widehat V}
\def\Vtilde{\widetilde V}
\def\wbar{\skew3\bar w}
\def\what{\skew3\widehat w}
\def\wtilde{\skew3\widetilde w}
\def\Wbar{\skew3\bar W}
\def\What{\widehat W}
\def\Wtilde{\widetilde W}
\def\xbar{\skew{2.8}\bar x}
\def\xhat{\skew{2.8}\widehat x}
\def\xtilde{\skew3\widetilde x}
\def\Xhat{\widehat X}
\def\ybar{\skew3\bar y}
\def\yhat{\skew3\widehat y}
\def\ytilde{\skew3\widetilde y}
\def\Ytilde{\widetilde Y}
\def\Ybar{\skew2\bar Y}
\def\Yhat{\widehat Y}
\def\zbar{\skew{2.8}\bar z}
\def\zhat{\skew{2.8}\widehat z}
\def\ztilde{\skew{2.8}\widetilde z}
\def\Zbar{\skew5\bar Z}
\def\Zhat{\widehat Z}
\def\Ztilde{\widetilde Z}
\def\MINOS{{\small MINOS}}
\def\NPSOL{{\small NPSOL}}
\def\QPSOL{{\small QPSOL}}
\def\LUSOL{{\small LUSOL}}
\def\LSSOL{{\small LSSOL}}

 \def\v#1{\texttt{#1}}
 \def\ttt#1{\texttt{#1}}
 \def\mv#1{\mathtt{#1}}
 \def\mtt#1{\mathtt{#1}}

\def\fr{_{\scriptscriptstyle{\it FR}}}
\def\fx{_{\scriptscriptstyle{\it FX}}}
\def\IN{{\bf (Input)}\hskip 1em\ignorespaces}
\def\OUT{{\bf (Output)}\hskip 1em\ignorespaces}

 
\newenvironment{parameters}[2]%
	{%
	 \begin{list}%
	  {$\bullet$}%
	  {\itemsep 2pt plus 1pt minus 1pt
	        \topsep 2pt plus 1pt minus 1pt 
	        \settowidth{\labelwidth}{#1}
	        \settowidth{\leftmargin}{#1}
		\addtolength{\leftmargin}{\labelsep}}
	   #2}%
	{\end{list}}

  \def\parm#1{\item[\tt#1\hfill]}
  \def\Item#1{\item[#1\ \ ]}
  \def\mitem#1{\item[$#1$\hfill]}
  \def\mitemwidth{\hbox{$<0$\ \ }}
  \def\optskip {\bigskip\medskip}
  \def\optsep  {5pt}
  \def\optwidth{2.4in}
  \def\wideoptwidth{3in}

  \def\option #1 #2 #3{
	\optskip
	\hbox to \textwidth
   	  {\hbox to \optwidth{\tt #1\hfil}$#2$\hfil Default = $#3$}
	\nobreak\vspace{\optsep}\noindent\ignorespaces}

  \def\wideoption #1 #2 #3{
	\optskip
	\hbox to \textwidth
   	  {\hbox to \wideoptwidth{\tt #1\hfil}$#2$\hfil Default = $#3$}
	\nobreak\vspace{\optsep}\noindent\ignorespaces}

  \def\optionb #1 #2 #3{
	\par\vspace{-\optsep}
	\hbox to \textwidth
	  {\hbox to \optwidth{\tt #1\hfil}$#2$\hfil Default = $#3$}
	\nobreak\vspace{\optsep}\noindent\ignorespaces}

  \def\wideoptionb #1 #2 #3{
	\par\vspace{-\optsep}
	\hbox to \textwidth
	  {\hbox to \wideoptwidth{\tt #1\hfil}$#2$\hfil Default = $#3$}
	\nobreak\vspace{\optsep}\noindent\ignorespaces}

  \def\optionc #1 #2 #3{
	\par\vspace{-\optsep}
	\hbox to \textwidth
	  {\hbox to \optwidth{\tt #1\hfil}$#2$\hfil #3}
	\nobreak\vspace{\optsep}\noindent\ignorespaces}

  \def\Option #1 #2{
	\optskip
	\hbox to \textwidth
	  {\hbox to \optwidth{\tt #1\hss}{\tt #2}\hfil Default}
	\nobreak\vspace{\optsep}\noindent\ignorespaces}

  \def\OOption #1 #2{
	\optskip
	\hbox to \textwidth
	  {\hbox to \optwidth{\tt #1\hss}{\tt #2}\hfil}
	\nobreak\vspace{\optsep}\noindent\ignorespaces}

  \def\Optionb #1 #2{
	\par\vspace{-\optsep}
	\hbox to \textwidth
	  {\hbox to \optwidth{\tt #1\hss}{\tt #2}\hfil}
	\nobreak\vspace{\optsep}\noindent\ignorespaces}

  \def\Optionc #1 #2{
	\optskip
	\hbox to \textwidth
	  {\hbox to \optwidth{\tt #1\hss}{\tt #2}\hfil}
	\nobreak\noindent\ignorespaces}

  \def\Heading #1 #2{%
	\hbox to \textwidth
	  {\hbox to \optwidth{\tt #1\hss}{\tt #2}\hfil Default}
	\nobreak\vspace{\optsep}\noindent\ignorespaces}

\def\figcap{\list{}{\listparindent 4em
 \itemindent\listparindent
 \rightmargin\leftmargin\parsep 0pt plus 1pt
 \topsep 4pt plus 2pt minus 4pt}\item[]}
\let\endfigcap=\endlist


 \def\Indexsf#1{\index{#1@\textsf{#1}}}
 \def\Indextt#1{\index{#1@\texttt{#1}}}

 \def\ka#1{k\mskip -0.75 mu\mbox{\it#1}}
 \def\kb#1{k\mskip -0.9 mu\mbox{\scriptsize\it#1}}

 \def\thefootnote{\fnsymbol{footnote}}
 \def\strut{\rule[-1.25ex]{0pt}{4ex}}%
 \def\strutl{\rule[-1.25ex]{0pt}{3ex}}%
 \def\strutu{\rule{0pt}{3ex}}%

 \def\Jac{F'}
 \def\Hessq{H}
 \def\gradq{g}

\def\problem#1#2#3#4{\fbox
   {\begin{tabular*}{0.84\textwidth}
    {@{}l@{\extracolsep{\fill}}l@{\extracolsep{6pt}}l@{\extracolsep{\fill}}c@{}}
      #1 & $\minimize{#2}$ & $#3$ & $ $ \\[5pt]
         & $\subject$      & $#4$ & $ $
    \end{tabular*}}}

\def\MPS     {{\small MPS}}
\def\BASIS   {{\small BASIS}}
\def\NEWBASIS{{\small NEW BASIS}}
\def\BACKUPBASIS{{\small BACKUP BASIS}}
\def\OLDBASIS{{\small OLD BASIS}}
\def\SPECS   {{\small SPECS}}
\def\INSERT  {{\small INSERT}}
\def\LOAD    {{\small LOAD}}
\def\DUMP    {{\small DUMP}}
\def\PUNCH   {{\small PUNCH}}
\def\PRINT   {{\small PRINT}}
\def\SUMMARY {{\small SUMMARY}}
\def\SOLUTION{{\small SOLUTION}}
\def\EXIT{{\small EXIT}}
\def\EXPAND{{\small EXPAND}}
\def\MINOS {{\small MINOS}}
\def\LUSOL {{\small LUSOL}}
\def\LPOPT {{\small LPOPT}}
\def\QPOPT {{\small QPOPT}}
\def\NPSOL {{\small NPSOL}}
\def\NPOPT {{\small NPOPT}}
\def\SQOPT {{\small SQOPT}}

\def\PRICE {{\small PRICE}}
\def\CRASH {{\small CRASH}}
\def\UNBOUNDED {{\small UNBOUNDED}}
\def\INFEASIBLE{{\small INFEASIBLE}}

\def\usrfun{{\tt usrfun}}
\def\usrini{{\tt usrini}}
\def\snInit{{\tt snInit}}
\def\snSpec{{\tt snSpec}}
\def\snset {{\tt snset}}
\def\snsetr{{\tt snsetr}}
\def\snseti{{\tt snseti}}
\def\sngetc{{\tt sngetc}}
\def\sngetr{{\tt sngetr}}
\def\sngeti{{\tt sngeti}}
\def\snMem {{\tt snMem}}
\def\snopt {{\tt snopt}}
\def\sqopt {{\tt sqopt}}
\def\GNUmakefile{\texttt{GNUmakefile}}

\def\Hline{\par\noindent\rule{\textwidth}{0.6pt}}

\def\NP#1{NP$(#1)$}
\def\QP#1{QP$(#1)$}
\def\Phase#1#2{Phase#1$(#2)$}

\def\SNADIOPT{\textsf{SnadiOpt}}
\def\ADIFOR{\textsf{ADIFOR}}
\def\C{\textsf{C}}
\def\FORTRAN{\textsf{Fortran}}
\def\SNOPT{\textsf{Snopt}}
\def\Snopt{\textsf{Snopt}}
\def\GNU{\textsf{GNU}}
\def\PERL{\textsf{Perl}}
\def\NEOS{\textsf{NEOS}}
\def\AMPL{\textsf{AMPL}}
\def\GAMS{\textsf{GAMS}}
\def\UNIX{\textsf{Unix}}
\def\MAKE{\textsf{Make}}
\def\BLAS{\textsf{BLAS}}
\pagenumbering{roman}
\setcounter{page}{1}
\thispagestyle{empty}
\begin{center}
\vspace*{-1in}
Argonne National Laboratory \\
9700 South Cass Avenue\\
Argonne, IL 60439

\vspace{.2in}
\rule{1.5in}{.01in}\\ [1ex]
ANL/MCS-TM-245 \\
\rule{1.5in}{.01in}

\vspace{1in}
{\large\bf Users Guide for SnadiOpt:  A Package Adding\\ [1ex]
Automatic Differentiation to Snopt}\footnote{This
work was supported by the Mathematical,
Information, and Computational Sciences Division subprogram of the
Office of Advanced Scientific Computing Research, U.S. Department of
Energy, under Contract W-31-109-Eng-38, and by National Science
Foundation Grant CCR-95-27151.}


\vspace{.2in}
by \\ [3ex]

{\large\it E. Michael Gertz}\\
{\small Mathematics and Computer Science Division}\\
{\small Argonne National Laboratory}\\
{\small Argonne, Illinois 60439}

\vspace{.2in}

{\large\it Philip E. Gill and Julia Muetherig}\\ 
{\small Department of Mathematics}\\
{\small University of California, San Diego}\\
{\small La Jolla, California 92093-0112}

\thispagestyle{empty}

\vspace{1in}
Mathematics and Computer Science Division

\bigskip

Technical Memorandum No. 245

\vspace{.5in}
January 2001
\end{center}

\vfill
\noindent
{\footnotesize
\begin{tabular}{lll}  
   gertz@mcs.anl.gov
&  pgill@ucsd.edu
&  jmueth@ucsd.edu
\\ http://www.mcs.anl.gov/$\sim$gertz/
&  http://www.scicomp.ucsd.edu/$\sim$peg/
&  http://www.scicomp.ucsd.edu/$\sim$julia/
\end{tabular}
}

\newpage
\pagenumbering{roman}
\setcounter{page}{2}
\hbox{}
\newpage
\noindent
Argonne National Laboratory, with facilities in the states of Illinois
and Idaho, is owned by the United States Government and operated by The
University of Chicago under the provisions of a contract with the
Department of Energy.

\vspace{2in}

\begin{center}
{\bf DISCLAIMER}
\end{center}

\noindent
This
report was prepared as an account of work sponsored by an agency of the United States Government.  Neither the United States
Government nor any agency thereof, nor The University of Chicago, nor any of
their employees or officers, makes any warranty, express or implied, or assumes any legal liability or
responsibility for the accuracy, completeness, or usefulness of any
information, apparatus, product, or process disclosed, or represents that its use
would not infringe privately-owned rights.
Reference herein to any specific commercial product, process, or service by trade name,
trademark, manufacturer, or otherwise, does not necessarily constitute
or imply its endorsement, recommendation, or favoring by the United
States Government or any agency thereof.  The views and opinions of document
authors expressed herein do not necessarily state or reflect those of the
United States Government or any agency thereof, Argonne National
Laboratory, or The University of Chicago.
\newpage
\hbox{}
\newpage
  \markboth{Users Guide for {\textsf SnadiOpt}}{}
  \pagestyle{plain}
  \tableofcontents
\newpage
  \newpage

\newpage

  \pagestyle{headings}
\pagenumbering{arabic}
  \setcounter{page}{1}

\vspace{1in}

\begin{center}
{\bf Users Guide for {\textsf SnadiOpt}: A Package Adding\\ [1ex]
Automatic Differentiation to {\textsf Snopt}} \\[2ex]
by \\ [2ex]
{\it E. Michael Gertz, Philip E. Gill, and Julia Muetherig}
\end{center}

\bigskip

\addcontentsline{toc}{section}{Abstract}
\begin{abstract}
 \SNADIOPT\ is a package \index{SnadiOpt@{\textsf{SnadiOpt} package}}
that supports the use of the automatic
differentiation package \ADIFOR\ with the optimization package \SNOPT.

 \SNOPT\ is a general-purpose system for solving optimization problems
with many variables and constraints.  It minimizes a linear or
nonlinear function subject to bounds on the variables and sparse
linear or nonlinear constraints.  It is suitable for large-scale
linear and quadratic programming and for linearly constrained
optimization, as well as for general nonlinear programs.

  The method used by \SNOPT\ requires the first derivatives of the
objective and constraint functions to be available.  The \SNADIOPT\
package allows users to avoid the time-consuming and error-prone
process of evaluating and coding these derivatives.  Given \FORTRAN\
code for evaluating only the \EM{values} of the objective and
constraints, \SNADIOPT\ automatically generates the code for
evaluating the derivatives and builds the relevant \SNOPT\ input files
and sparse data structures. 
\index{Snopt@\textsf{Snopt}}

\bigskip

Keywords: Large-scale nonlinear programming, constrained optimization, SQP
          methods, automatic differentiation, \FORTRAN\  software.
\index{constrained optimization|see{nonlinear constrained optimization}}
\index{SQP methods} \index{differentiation!automatic}
\index{Fortran@{\textsf{Fortran} compiler}}
\end{abstract}
\newpage

\def\Obj{F_{\mbox{\it\small obj}}}
\def\xlow{l}
\def\xupp{u}
\def\Flow{L}
\def\Fupp{U}

 \section{Introduction}  

This is the users guide for \SNADIOPT,
\index{SnadiOpt@{\textsf{SnadiOpt} package}} a package that adds the automatic
differentiation capability \index{differentiation!automatic} to the
nonlinear optimization 
\index{nonlinear constrained optimization}%
\index{nonlinear programming}%
package \SNOPT\ \cite{GMS97}.  
\index{Snopt@\textsf{Snopt}!nonlinear optimization package} 
\SNADIOPT\ uses the
source-to-source automatic differentiation package \ADIFOR\
\index{ADIFOR@\textsf{ADIFOR}} to perform the
differentiation.

 \subsection{Problem Types} 

 \SNOPT\ is a collection of \FORTRAN~77 subroutines for solving a
\EM{nonlinear programming problem} assumed to be stated in the 
following form:
$$
\fbox{%
\begin{tabular*}{0.84\textwidth}
    {@{}l@{\extracolsep{\fill}}l@{\extracolsep{6pt}}l@{\extracolsep{\fill}}c@{}}
      NP & $\disp\min_{x}$ (or max)  & $\Obj(x)$                        & $ $ \\[5pt]
         & $\subject$                & $\xlow \le   x  \le \xupp, \qquad
                                        \Flow \le F(x) \le \Fupp,$ & $ $
\end{tabular*}}
$$
where $\xupp$, $\Fupp$, $\xlow$, and $\Flow$ are constant vectors of
upper and lower bounds, $F(x)$ is a vector of smooth linear and
nonlinear problem functions, and $\Obj(x)$ denotes the component of
$F$ to be minimized or maximized.
\index{bounds}
\index{F@$F(x)$}
\index{F1@$\Obj(x)$}
\index{l@$\xlow$}
\index{u@$\xupp$}
\index{L@$\Flow$}
\index{U@$\Fupp$}

Note that upper and lower bounds are specified for all variables and
constraints.  \index{constraints} This form allows full generality in
specifying various types of constraint.  Special values are used to
indicate absent bounds ($l_j = -\infty$ or $u_j = +\infty$ for
appropriate $j$).  Free variables and free constraints (``free rows'')
are ones that have both bounds infinite.  Similarly, fixed variables
have $\xlow_j=\xupp_j$, and equality constraints have
$\Flow_j=\Fupp_j$

The method used by \SNOPT\ requires that the elements
$J_{ij}(x)=\partial F_i(x)/\partial x_j$ of the Jacobian matrix
\index{Jacobian matrix (J)} of first derivatives be known at any point
$x$.  In practice it is often inconvenient or impossible to code the
derivatives, and so \SNOPT\ allows the user to code as many
derivatives as is convenient.  \SNOPT\ then estimates unknown
derivatives by finite differences, by making a call to $F$ for each
variable $x_j$ whose partial derivatives need to be estimated.
However, finite differences reduce the reliability of the
optimization algorithm and can be expensive if there are many
such variables $x_j$.  The \SNADIOPT\ package allows the user to avoid
the time-consuming and error-prone process of evaluating and coding
derivatives without the need for \SNOPT\ to compute
finite differences.

  Often, an element $J_{ij}$ is constant,
\index{Jacobian matrix (J)!constant elements} which implies that variable
$x_j$ occurs only linearly in the problem function $F_j(x)$.  If a
significant number of these constant elements are zero, then $J$ is
known as a sparse matrix, and \SNOPT\ uses a sparse
matrix format to store only the nonzero elements of $J$.
\index{Jacobian matrix (J)!sparse matrix format} \SNADIOPT\
automatically identifies constant and zero Jacobian elements by using a
scheme that evaluates the Jacobian at a number of points close to the
starting point (see Section~\ref{automatic-differentiation}).  Given
\FORTRAN\ code for evaluating only $F(x)$, \SNADIOPT\ automatically
generates code for evaluating $J$ and builds the relevant \SNOPT\
input files and sparse data structures. 
\index{Snopt@\textsf{Snopt}!sparse optimization}
\index{SnadiOpt@{\textsf{SnadiOpt} package}!treatment of sparse
Jacobians}


\index{AD (automatic differentiation)|(}

\subsection{Why Automatic Differentiation?}

Writing code for the derivatives of $F(x)$ \index{derivatives of
\v{F}} is difficult, time consuming, and error prone, especially when
problems involve many variables and constraints. Automatic
differentiation (AD) tools\index{differentiation!automatic}, in this
case \ADIFOR~\cite{adifor, adifor2.0}, \index{ADIFOR@\textsf{ADIFOR}}
quickly provide correct and numerically accurate derivative functions
from the code used to evaluate the objective and constraint functions.

Prior to the wide availability of AD software and AD-based modeling
languages, numerical differentiation \index{differentiation!numerical}
was the only alternative to providing derivative code.  Unfortunately,
numerical differentiation is an inherently unstable process that
causes both theoretical and practical difficulties for nonlinear
solvers.  Numerical differentiation places a severe theoretical limit
on the accuracy of the solution that may be computed by an algorithm
(see, e.g., \cite[Chapter~8]{GMW81}). In practice, code that uses
numerical differentiation tends to need more iterations to find a
solution than does code that uses exact derivatives.  Furthermore, code
that uses numerical differentiation is typically less robust and will
fail to find solutions for problems that might have been solved if
analytic derivatives were supplied. Notwithstanding these
difficulties, numerical differentiation was often used to avoid the
high cost of hand-coding the exact
derivatives. \index{differentiation!numerical}

With modern AD tools, derivative code may be quickly obtained, leading
to a significant increase in user productivity---even on simple
problems.  Moreover, functions can now be differentiated that were
once considered too complex to be coded by hand.  An example 
is a function defined in terms of the output from an ordinary
differential equation solver. \index{ordinary differential equation solver}
\ADIFOR\ has been successfully applied to such functions.
\index{differentiation!automatic} \index{ADIFOR@\textsf{ADIFOR}}

Automatic differentiation \index{differentiation!automatic}
allows users to develop models quickly. This
increase in productivity makes optimization software a much more
useful tool for scientists, who often wish to experiment with
different objective functions and different sets of constraints.

\subsection{\ADIFOR}
\index{ADIFOR@\textsf{ADIFOR}}

\ADIFOR\ is a robust, mature automatic differentiation
\index{differentiation!automatic} tool developed through a
collaborative project between the Mathematics and Computer Science
Division at Argonne National Laboratory and the Center for Research on
Parallel Computation at Rice University.  \index{Rice University}
\index{Argonne National Laboratory} The package is a source-to-source
translator for functions written in \textsf{Fortran~77}%
\index{Fortran@{\textsf{Fortran} compiler}!fortran 77@\FORTRAN~77}.  
It is widely
available and runs on many popular platforms.  Moreover, the source
code for the \ADIFOR\ libraries
\index{ADIFOR@\textsf{ADIFOR}!libraries} (but not the translator) is
provided, making it possible to compile and run \ADIFOR-generated code
on most platforms.

\ADIFOR\index{ADIFOR@\textsf{ADIFOR}} implements the forward mode for
automatic differentiation \index{forward mode (for AD)} (although it
does make some use of the reverse mode internally). As a practical
matter, this means that \ADIFOR-generated code will tend to run most
quickly if the number of variables is not much larger than the number
of constraints.


 \subsection{Who Should Use This Package} 

It is usually wise to try automatic differentiation before attempting
to code derivatives by hand. \index{differentiation!automatic} Some
users may be more comfortable using modeling languages with an
automatic differentiation capability (see, e.g., \AMPL\ \cite{ampl}
\index{AMPL@\textsf{AMPL}} and \GAMS\ \cite{gams}
\index{GAMS@\textsf{GAMS}}).  \SNADIOPT\ is intended for those who
prefer to code in \FORTRAN\index{Fortran@{\textsf{Fortran}
compiler}}, or need to make use of existing \FORTRAN\ software.  Such
users should find that this package can 
provide derivative code quickly and efficiently.

Programmers who typically write in \C\ \index{C@{\textsf{C} compiler}}
or \textsf{Fortran~90} \index{Fortran@{\textsf{Fortran}
compiler}!fortran 90@\FORTRAN~90} might like to consider developing their models
in \textsf{Fortran~77} so that they may use this package to obtain
derivatives.  A \C\ programmer should be able to learn enough
\textsf{Fortran~77} to formulate simple to moderately complicated
models within a few hours. \C\ has many features that
\textsf{Fortran~77} \index{Fortran@{\textsf{Fortran}
compiler}!fortran 77@\FORTRAN~77} does not, but it is exactly those 
features that can make the automatic differentiation of \C\ code 
problematic.


\index{AD (automatic differentiation)|)}

 \subsection{How to Read This Manual} 

Section~\ref{sec-basic-usage} summarizes the four steps
needed to define and solve an optimization model using \SNADIOPT\ and
\SNOPT.  Section~\ref{automatic-differentiation} describes the main
features of the \SNADIOPT\ package and provides some background on the
mechanics of automatic differentiation (this section may be omitted on
first reading).  Section~\ref{user-supplied-subroutines} describes the
subroutines that must be provided by users to define their model.
Section~\ref{invoking} descibes the invocation of the \PERL\ script
\v{snadiopt.pl} that automatically generates all input files for
\ADIFOR.  This discussion includes detailed information on the various
files generated by \SNADIOPT\ (see Section~\ref{generated-files}).
Finally, Section~\ref{makefile} discusses the use of the \MAKE\
utility to automatically differentiate the problem files and build an
executable file ready for execution.

\index{SnadiOpt@{\textsf{SnadiOpt} package}!basic usage|(}

 \subsection{Basic Usage}\label{sec-basic-usage}   

In the simplest case, the \SNADIOPT\ package can be used to solve
problem NP in four steps.    \index{snadiopt.pl@\texttt{snadiopt.pl}!basic usage}

\begin{enumerate}
 \item[Step 1.] Construct a file \v{prob.f} containing \FORTRAN\
   subroutines \v{usrini} and \v{usrfun} (see
   Section~\ref{user-supplied-subroutines}).  Subroutine \v{usrini}
   initializes all data associated with the model, including the
   bounds $\xlow$, $\xupp$, $\Flow$, and $\Fupp$, and the component of
   $F$ that defines the objective function.  Subroutine \v{usrfun}
   defines the values of the problem functions $F$ at a given value of
   $x$.  A \FORTRAN\ main program is not required.  However,
   subroutine \v{usrini} must define the lengths of all arrays
   used to define the model.
   \index{usrfun@{\texttt{usrfun} (user-supplied subroutine)}}%
   \index{usrini@{\texttt{usrini} (user-supplied subroutine)}}%
   \index{prob@{\texttt{prob} files}!prob.f@\v{prob.f} (user supplied)}%
   \index{prob@{\texttt{prob} files}!fortran source@\FORTRAN\ source files}
   \index{l@$\xlow$} \index{u@$\xupp$} \index{L@$\Flow$}%
   \index{U@$\Fupp$}%

 \item[Step 2.] Invoke the \PERL\ script \v{snadiopt.pl} to generate the
   input files needed by the automatic differentiation package
   \ADIFOR.  The syntax of the call is \\
   \v{\% snadiopt.pl -o prob prob.f}   \\
   (The symbol ``\v{\%}'' is the shell prompt and is not to be typed).  This
   generates a number of auxiliary files with the prefix \v{prob\_}
   (see Section~\ref{invoking}).
   \index{Perl@\textsf{Perl}}
   \index{snadiopt.pl@\texttt{snadiopt.pl}}

 \item[Step 3.] Build the executable file for the model \v{prob}.
   This step  uses \ADIFOR\ to generate  the differentiated
   subroutines needed for \SNOPT, compiles them and links them with
   the \SNOPT\ libraries.  All these tasks are performed by using the
   \MAKE\ utility, where the \v{makefile} is one of the files
   automatically generated by \v{snadiopt.pl} (see
   Section~\ref{makefile}).  To start the build process, type\\
   \v{\% make prob}

 \item[Step 4.] Solve the optimization problem by typing \\
   \v{\% ./prob} \\
   Any output from the run will be written to the files defined in the
  subroutine \v{usrini}. 
  \index{prob@{\texttt{prob} files}!prob (executable)@\v{prob} (executable)}

\end{enumerate}

\index{SnadiOpt@{\textsf{SnadiOpt} package}!basic usage|)}

 \subsection{Additional Resources} 

All users should read the \SNOPT\ users guide \cite{GMS97c},
\index{Snopt@\textsf{Snopt}!users guide} which details the many user options
available in \SNOPT\ that may be set by providing a ``specs'' file.
\index{Snopt@\textsf{Snopt}!specs file} The users guide describes the
algorithm and its output and answers many questions about the performance
of the optimizer on a particular model.

 \SNADIOPT\ tries to insulate the user from the details of invoking \ADIFOR,
but users may wish to read the \ADIFOR\ users guide.
\index{ADIFOR@\textsf{ADIFOR}!users guide} In particular, while \ADIFOR\ is
very robust, it is possible to write \FORTRAN\ code that fools \ADIFOR, and the
manual will explain how to avoid this pitfall.

For general information on optimization, we recommend that users
explore the \NEOS\  guide on the Web: \v{http://www.mcs.anl.gov/otc/Guide/index.html}.
\index{optimization Web page} \index{NEOS@\textsf{NEOS}}

More information about automatic differentiation in general, and
\ADIFOR\ in particular, may be found on the Argonne National Laboratory
automatic differentiation Web page: \v{http://www.mcs.anl.gov/autodiff}.
\index{ADIFOR@\textsf{ADIFOR}!Web page}

Users with complicated functions for which the automatically
differentiated code appears to be unacceptably slow can often
accelerate their code by refactoring it. Several technical reports on
the automatic differentiation page describe how to do this.

The authors of this package also maintain Web pages. Philip Gill's
page\footnote{\texttt{http://scicomp.ucsd.edu/\~{}peg/}} has links to published
papers and technical reports on \SNOPT.  Michael Gertz maintains a
Web page\footnote{\texttt{http://www.mcs.anl.gov/\~{}gertz/}} at Argonne National Laboratory.
\index{Snopt@\textsf{Snopt}!Web page}
\index{ADIFOR@\textsf{ADIFOR}!Web page}

\index{constraint functions|see{$F$ vector}}
\index{AD (automatic differentiation)|see{differentiation}}
\index{objective function|see{\v{ObjRow}}}


 \section{Automatic Differentiation}  \label{automatic-differentiation}
 \index{differentiation!automatic}

Automatic differentiation is the process of producing code that
evaluates the derivatives of a function from code that evaluates the
function itself. It is closely related to symbolic differentiation but
\index{differentiation!symbolic}
differs from it in important ways. Symbolic differentiation takes the
mathematical expression for a function and produces another expression
that represents the derivative of that function. Unlike
symbolic differentiation packages, automatic differentiation packages:
\begin{itemize}
 \item understand programming concepts such as loops, branches and
   subroutines  and 

 \item use intermediate quantities and the chain rule to avoid
   potentially exponential growth in the size of the resulting code.
\end{itemize}

\ADIFOR\ is a source-to-source translator:
\index{ADIFOR@\textsf{ADIFOR}!input/output} it takes
as its input a function expressed as a \textsf{Fortran~77} subroutine
and generates \FORTRAN\ code that computes the derivatives
of the dependent variables with respect to the independent variables.
Suppose $F:\Re^n \to \Re^m$ is a function and
\begin{verbatim}
       subroutine func  ( x, n, F, m )
       integer            m, n
       double precision   x(n), F(m)
\end{verbatim}
is a \FORTRAN\ subroutine that will compute the value of $F$ at any
given $x$.  Let $J(x) = F'(x)$ be the Jacobian of $F$.
\index{Jacobian matrix (J)} \index{F@{\texttt{F} (array of problem 
function values)}}
\ADIFOR\ will produce code that computes $J(x) S$ for any $n \times p$
matrix $S$ with $p \le n$. Thus, if $S=I$, \ADIFOR\ will
compute the Jacobian itself.

In many optimization models, certain terms that occur in $F$ will be 
linear and will result in constant elements in $J(x)$.  \SNADIOPT\ 
\index{SnadiOpt@{\textsf{SnadiOpt} package}} does not require that 
these elements occur in a particular part of $J(x)$, but for the sake 
of discussion, let us assume that $J(x)$ has the following structure, 
\index{Jacobian matrix (J)}
$$
    J(x) = \mtx{cc}{ N_{11} & L_{12} \\
                     L_{21} & L_{22} },
$$
where the elements of $L_{12}$, $L_{21}$, and $L_{22}$ are constant
and the elements of $N_{11}$ may or may not be constant, but all rows
and columns of $N_{11}$ contain at least one nonconstant element. Any
Jacobian may be transformed to a matrix with this structure by
permuting the constraints (rows) and variables (columns).  \SNOPT\ is
designed to exploit the constant elements in $J$. For instance, the
constraints corresponding to \index{Snopt@\textsf{Snopt}!input of the
Jacobian} $\tmat{L_{21}}{L_{22}}$ are linear, and \SNOPT\ will
maintain feasibility with respect to the linear constraints. Because
the \ADIFOR\ generated code computes $J(x)S$, \SNADIOPT\ is able to
choose an $S$ in a manner that avoids the need to reassign $L_{12}$,
$L_{21}$, and $L_{22}$ every time $J$ is required.

 \SNOPT\ \index{Snopt@\textsf{Snopt}!input of the Jacobian} is
designed to solve problems with sparse derivatives.  These are
problems for which many of the elements of $J(x)$ 
\index{Jacobian matrix (J)}
are identically zero.  \SNADIOPT\ \index{sparse Jacobian}
\index{SnadiOpt@{\textsf{SnadiOpt} package}} determines the sparsity
pattern for the Jacobian and identifies the constant elements
automatically. To make this determination, \SNADIOPT\ computes the
value of $J(x)$ at two random perturbations of a user-supplied
initial point $x_0$.  \index{initial point $x_0$, calculation of sparse
Jacobian} If an element of the Jacobian is the same at both points,
then it is taken to be constant. If it is zero at both points, it is
taken to be identically zero.  The random points are not chosen close
together, so the heuristic will correctly classify the Jacobian
elements in the vast majority of cases.  \SNOPT\
\index{Snopt@\textsf{Snopt}!input of the Jacobian} validates the
computed derivatives, linearity pattern, and sparsity pattern%
\index{Jacobian matrix (J)!linearity pattern}%
\index{Jacobian matrix (J)!sparsity pattern}
at the point $x_0$, by comparing the supplied
values to values computed using numerical differentiation.  This
additional test at a third point makes it unlikely that an
incorrect sparsity or linearity pattern will be used.
\index{differentiation!numerical}

 Of course, it is possible to fool this heuristic. \SNADIOPT\
\index{SnadiOpt@{\textsf{SnadiOpt} package}} cannot deal well with functions for
which the sparsity pattern or linearity pattern in a (relatively
large) region around $x$ is not representative of the sparsity or
linearity pattern of the function as a whole. Computing a sparsity
pattern for such a function would require significant additional user
intervention. Because we are uncertain of the demand to
minimize such functions, we have opted for the simpler user
interface. We welcome examples of real-word optimization models
that fall into this category.

Once \SNADIOPT\ \index{SnadiOpt@{\textsf{SnadiOpt} package}} has
computed the sparsity and linearity pattern\index{linearity pattern}
and the appropriate $S$ to minimize recomputation of the derivatives
of linear elements, it calls \SNOPT\ \index{Snopt@\textsf{Snopt}!call
to \SNOPT} as a ``black-box'' optimization routine. This means that it
presents the optimization data to \SNOPT\ in the same format as a
hand-written routine for computing the derivatives. Users have full
access to all the options and features of \SNOPT\ and can link the
resulting code with their own code (subject, of course, to any
licensing restrictions.)

\index{derivatives of $F$|see{Jacobian matrix (J)}}



\section{User-Supplied Subroutines}  \label{user-supplied-subroutines}

In order to use \SNADIOPT, the user must provide the following 
\FORTRAN\ routines:
$$
\fbox{\begin{minipage}{0.98\textwidth}
\begin{description}
 \item[\usrfun] (\Sec\ref{sec-usrfun}) Defines the functions $F_i(x)$. \index{F@$F(x)$}

 \item[\usrini] (\Sec\ref{sec-usrini}) Defines the actual dimension of
        the problem and initializes all data needed by \SNOPT.  The
        workspace for \SNADIOPT\ is also assigned here.
\end{description}
\end{minipage}}
$$
The user routines \usrfun\ and \usrini\ have fixed parameter lists but
may have any convenient name. The names of the parameters may also be
chosen by the user.

\index{usrfun@{\texttt{usrfun} (user-supplied subroutine)}}
\index{usrini@{\texttt{usrini} (user-supplied subroutine)}}

 \subsection{The Function Definition Routine} \label{sec-usrfun}

The user must provide a subroutine that calculates the vector $F$ of
objective and constraint functions at a given point $x$. 
\index{usrfun@{\texttt{usrfun} (user-supplied subroutine)}!specification}
\begin{verbatim}
      subroutine usrfun( Status, mode, x, n, F, neF,
     &                   cu, lencu, iu, leniu, ru, lenru )
      integer            Status, mode, neF, n
      double precision   F(neF), x(n) 
      integer            lencu, leniu, lenru
      character*8        cu(lencu)
      integer            iu(leniu)
      double precision   ru(lenru)
\end{verbatim}
\index{Status@\texttt{Status}!argument of \v{usrfun}}
\index{x@{\texttt{x} (array of variables)}!argument of \v{usrfun}}
\index{F@{\texttt{F} (array of problem function values)}!argument of \v{usrfun}}
\index{scratch arrays!character, \v{cu(lencu)}!argument of \v{usrfun}}
\index{scratch arrays!integer, \v{iu(leniu)}!argument of \v{usrfun}}
\index{scratch arrays!real, \v{ru(lenru)}!argument of \v{usrfun}}
\index{mode@\texttt{mode}!argument of \v{usrfun}}
\index{cu@{\texttt{cu(lencu)} (user character scratch array)}!argument of \v{usrfun}}
\index{iu@{\texttt{iu(leniu)} (user integer scratch array)}!argument of \v{usrfun}}
\index{ru@{\texttt{ru(lenru)} (user double scratch array)}!argument of \v{usrfun}}
\subsection*{On entry:}
\begin{parameters}{\v{inform}}
  \parm{Status} indicates the first and last call to \usrfun.
    \index{Status@\texttt{Status}!description}

    If $\v{Status}=0$, there is nothing special about the current
    call to \v{usrfun}.

    If $\v{Status} = 1$, \SNOPT\ is calling the subroutine for the
    first time.  Some data may need to be input or computed and
    saved.

    If $\v{Status} \ge 2$, \SNOPT\ is calling the subroutine for the
    last time.  The user may wish to perform some additional computation
    using the final solution.

    If the nonlinear functions are expensive to evaluate, it may be
    desirable to do nothing on the last call, by including a statement of
    the form
\begin{verbatim}
      if (Status .ge. 2) return
\end{verbatim}
    at the start of the subroutine.

 \parm{x(n)} contains the point at which the problem functions are to be
   evaluated.
    \index{x@{\texttt{x} (array of variables)}!description}

 \parm{cu(lencu), iu(leniu), ru(lenru)} are character, integer and real scratch
    arrays.  These arrays may be used to store any information that needs to be
    saved between calls to \v{usrfun}.
     \index{scratch arrays!character, \v{cu(lencu)}!description}
     \index{scratch arrays!integer, \v{iu(leniu)}!description}
     \index{scratch arrays!real, \v{ru(lenru)}!description}
\end{parameters}

\subsection*{On exit:}

\begin{parameters}{\v{inform}}
 \parm{F(neF)} holds the values of the objective and constraint
  functions computed at \v{x}. The objective is component 
  \v{F(ObjRow)},
  as defined in \usrini.
  \index{F@{\texttt{F} (array of problem function values)}!description}
  \index{ObjRow@{\texttt{ObjRow} (objective row)}!component of $F$}
  \index{F@$F(x)$}\index{F1@$\Obj(x)$}

  \parm{mode} is used to communicate between \SNOPT\ and the user.  If
  the user is unwilling or unable to evaluate the function at the current
  point then $\v{mode}$ should be set to $-1$. \SNOPT\ will try to provide an
  alternative point at which to evaluate the function.
  \index{mode@\texttt{mode}!description}

  If for some reason the user wishes to terminate solution of the current
  problem, \v{mode} should be set to a negative value (other than $-1$).

 \parm{cu(lencu), iu(leniu), ru(lenru)} are character, integer, and double
  precision arrays in which the user may store information between 
  calls to \usrfun.
  \index{cu@{\texttt{cu(lencu)} (user character scratch array)}!description}
  \index{iu@{\texttt{iu(leniu)} (user integer scratch array)}!description}
  \index{ru@{\texttt{ru(lenru)} (user double scratch array)}!description}
\end{parameters}


 \subsection{The Initialization Routine}  \label{sec-usrini}

\index{usrini@{\texttt{usrini} (user-supplied subroutine)}!specification}

 Subroutine \v{usrini} is used to initialize quantities associated 
with the problem.  It is called once before \SNOPT.
\begin{verbatim}
      subroutine usrini( ObjAdd, ObjRow, Prob, 
     &   x, xlow, xupp, xstate, Names,
     &   Fmul, Flow, Fupp, Fstate, FNames,
     &   iSpecs, iPrint, iSumm, iErr,     
     &   cu,  iu,  ru, 
     &   cw,  iw,  rw )
\end{verbatim}

Each argument is fully described below.  Many of the
arguments are arrays (e.g., \v{x} is the vector containing an initial
guess at the solution).  However, we emphasize that \EM{the
lengths of the array arguments do not appear in the argument list}. 
The user
must declare all arrays to be of fixed length at the head of the
subroutine \v{usrini}.  These declarations are used by the \PERL\  script
\v{snadiopt.pl} \index{Perl@\textsf{Perl}} \index{snadiopt.pl@\texttt{snadiopt.pl}}
to automatically construct a main program that calls
\SNOPT\  with appropriately dimensioned arrays.  For example, a
typical definition of the variables at the head of \v{usrini} is as
follows.
\begin{verbatim}
      integer            n, neF
      integer            nName, nFnames
      integer            lencw, leniw, lenrw, lencu, leniu, lenru
      parameter        ( lencw = 501, leniw = 10000, lenrw = 20000 )
      parameter        ( lencu =   1, leniu =     1, lenru =     1 )
      parameter        ( n     =   5, neF   =     6                )
      parameter        ( nName =   1, nFnames =   1                )
      character*8        Names(nName), FNames(nFnames)
      double precision   x(n), xlow(n), xupp(n)
      double precision   Flow(neF), Fupp(neF), Fmul(neF)
      integer            xstate(n),  Fstate(neF)
      integer            iu(leniu), iw(leniw)
      double precision   ru(lenru), rw(lenrw)
      character*8        cu(lencu), cw(lencw)
\end{verbatim}
\index{ObjAdd@{\texttt{Objadd} (objective additive constant)}!argument of \v{usrini}}
\index{ObjRow@{\texttt{ObjRow} (objective row)}!argument of \v{usrini}}
\index{Prob@{\texttt{Prob} (model name)}!argument of \v{usrini}}
\index{x@{\texttt{x} (array of variables)}!argument of \v{usrini}}
\index{xlow@{\texttt{xlow} (lower bounds on $x$)}!argument of \v{usrini}}
\index{xupp@{\texttt{xupp} (upper bounds on $x$)}!argument of \v{usrini}}
\index{xstate@{\texttt{xstate} (status of bounds on $x$)}!argument of \v{usrini}}
\index{Names@{\v{Names} (variable names)}!argument of \v{usrini}}
\index{Fmul@{\texttt{Fmul} (array of multipliers)}!argument of \v{usrini}}
\index{Flow@{\texttt{Flow} (lower bounds on $F$)}!argument of \v{usrini}}
\index{Fupp@{\texttt{Fupp} (upper bounds on $F$)}!argument of \v{usrini}}
\index{FNames@{\texttt{FNames} (names of $F$)}!argument of \v{usrini}}
\index{iSpecs@{\texttt{iSpecs} (specs file descriptor)}!argument of \v{usrini}}
\index{iPrint@{\texttt{iPrint} (print file descriptor)}!argument of \v{usrini}}
\index{iSumm@{\texttt{iSumm} (summary file descriptor)}!argument of \v{usrini}}
\index{iErr@{\texttt{iErr} (diagnostic file descriptor)}!argument of \v{usrini}}
\index{cu@{\texttt{cu(lencu)} (user character scratch array)}!argument of \v{usrini}}
\index{iu@{\texttt{iu(leniu)} (user integer scratch array)}!argument of \v{usrini}}
\index{ru@{\texttt{ru(lenru)} (user double scratch array)}!argument of \v{usrini}}
\index{cw@{\texttt{cw(lencw)} (\SNOPT\ character work array)}!argument of \v{usrini}}
\index{iw@{\texttt{iw(leniw)} (\SNOPT\ integer work array)}!argument of \v{usrini}}
\index{rw@{\texttt{rw(lenrw)} (\SNOPT\ double work array)}!argument of \v{usrini}}
\index{scratch arrays!character, \v{cu(lencu)}!argument of \v{usrini}}
\index{scratch arrays!integer, \v{iu(leniu)}!argument of \v{usrini}}
\index{scratch arrays!real, \v{ru(lenru)}!argument of \v{usrini}}
\index{work arrays}
The names of the arguments for \usrini\ are unimportant, but the
\EM{position} of each argument is significant.  For example, if the 
user
prefers to call the vector of variables ``\v{vars}'' and declares the
fourth argument of \usrini\ to be
\begin{verbatim}
      parameter         (maxvars = 5)
      double precision   vars(maxvars)
\end{verbatim}
then  \v{snadiopt.pl} will parse this declaration and 
include 
\begin{verbatim}
      parameter         (n = 5)
      double precision   x(n)
\end{verbatim}
in the automatically generated calling routine.  \SNADIOPT\ can recognize \FORTRAN\ 
\index{Fortran@{\textsf{Fortran} compiler}!code recognized by \SNADIOPT}
style parameters and numbers but cannot read more complicated
expressions. For instance, declaring \v{Fmul} as ``\v{double precision
Fmul(n+1)}'' will definitely confuse it.

Below, we describe each of the arguments of \usrini.  In many cases,
these arguments are assigned a default value in the automatically
generated calling program.  If the user wishes to use the default value of an
argument, then it should not be altered in \usrini.  The symbol ``$\infty$''
denotes the value of the \SNOPT\ optional parameter \v{Infinite
bound}, which has default value $10^{20}$. \index{Infinite bound}

\subsection*{Parameters:}
\begin{parameters}{\v{inform}}
 \parm{ObjAdd} is a double precision constant that is added to the
  objective function for printing purposes. \v{ObjAdd} does not affect
  the minimizer found.
  \index{ObjAdd@{\texttt{Objadd} (objective additive constant)}!description}

  Default value: $\v{ObjAdd} = 0.0$.
  \index{ObjAdd@{\texttt{Objadd} (objective additive constant)}!default value}

 \parm{ObjRow} is an integer defining the component of $F(x)$ to be
  used as the objective function $\Obj(x)$.  If $\v{ObjRow} = 0$,
  then \SNOPT\ finds a point $x$ that satisfies the
  constraints $\xlow \le x \le \xupp$, and $\Flow \le F(x) \le \Fupp$,
  \index{ObjRow@{\texttt{ObjRow} (objective row)}!description} 
  \index{F1@$\Obj(x)$}

  Default value: $\v{ObjRow} = 1$.
  \index{ObjRow@{\texttt{ObjRow} (objective row)}!default value} 

 \parm{Prob} is an eight-character name for this model.
    \index{Prob@{\texttt{Prob} (model name)}!description} 
 
  Default value: The name of the executable, truncated to eight characters.
    \index{Prob@{\texttt{Prob} (model name)}!default value}
 
 \parm{x} is a double precision array containing the point at which
  \SNOPT\ will start searching for a minimizer.
  \index{x@{\texttt{x} (array of variables)}!description}

  Default value: $\v{x}(j)= 0.0$.
  \index{x@{\texttt{x} (array of variables)}!default values}
 
  \parm{xlow, xupp} are double precision arrays containing the lower and upper
  bounds $\xlow$ and $\xupp$ such that $\xlow \le x \le \xupp$.  By default,
  \v{xlow} and \v{xupp} are assumed to be infinite (i.e., the value of
  $x$ is not restricted).
  \index{xlow@{\texttt{xlow} (lower bounds on $x$)}!description}
  \index{xupp@{\texttt{xupp} (upper bounds on $x$)}!description}
  \index{l@$\xlow$}\index{u@$\xupp$}

  Default values: $\v{xlow}(j)= -\infty$,  $\v{xupp}(j)= +\infty$.
  \index{xlow@{\texttt{xlow} (lower bounds on $x$)}!default values}
  \index{xupp@{\texttt{xupp} (upper bounds on $x$)}!default values}

 \parm{xstate} defines the initial state for each variable $x$.
  One may set $\v{xstate}(j) = 0$, $\v{x}(j) = 0.0$ for all $j=1:n$.  All
  variables will be eligible for the initial basis.
  \index{xstate@{\texttt{xstate} (status of bounds on $x$)}!description}
        
  Less trivially, to say that the optimal value of variable $j$ will
  probably be equal to one of its bounds, set $\v{xstate}(j) = 4$ and
  $\v{x}(j) = \v{xlow}(j)$ or $\v{xstate}(j) = 5$ and
  $\v{x}(j)=\v{xupp}(j)$ as appropriate.

  Default value: $\v{xstate}(j)= 0$.
  \index{xstate@{\texttt{xstate} (status of bounds on $x$)}!default values}
        
 \parm{Names} is a character array of symbolic names for the components
  of \v{x}. Each name may have up to eight characters. If the user does not
  wish to supply symbolic names for the variables, \v{Names} should be 
  declared to be 
  be an array of length one.
  \index{Names@{\v{Names} (variable names)}!description} 
  
 \parm{Fmul} is a double precision array of estimates of the dual variables for
  the constraints $\Flow \le F(x) \le \Fupp$. (Dual variables are sometimes
  known as Lagrange multipliers or shadow prices.)  \v{Fmul(ObjRow)} corresponds
  to the objective and is ignored.
  \index{Fmul@{\texttt{Fmul} (array of multipliers)}!description}

  Default value: $\v{Fmul}(j)= 0.0$.
  \index{Fmul@{\texttt{Fmul} (array of multipliers)}!default values}
  
 \parm{Flow, Fupp} are double precision arrays containing the lower and upper
  bounds $\Flow$ and $\Fupp$ such that $\Flow \le F_i(x) \le \Fupp$.  The
  components \v{Flow(ObjRow)} and \v{Fupp(ObjRow)} corresponding to the
  objective is ignored.  For an equality constraint of the form $F_i(x) = c$,
  set $\v{Flow}(j)=\v{Fupp}(j) = c$.
  \index{Flow@{\texttt{Flow} (lower bounds on $F$)}!description}
  \index{Fupp@{\texttt{Fupp} (upper bounds on $F$)}!description}
  \index{L@$\Flow$}
  \index{U@$\Fupp$}
  
  Default values: $\v{Flow}(j)= -\infty$ and $\v{Fupp}(j)= +\infty$.
  \index{Flow@{\texttt{Flow} (lower bounds on $F$)}!default values}
  \index{ObjRow@{\texttt{ObjRow} (objective row)}!bounds ignored}

 \parm{FNames} is a character array of symbolic names for the
  constraints.  Each name may consist of up to eight characters. If 
  the user
  does not wish to supply names for the constraints, \v{FNames} 
  should be declared to be
  an array of length one.
  \index{FNames@{\texttt{FNames} (names of $F$)}!description}
  
 \parm{iSpecs} is an open, readable \FORTRAN\ file descriptor pointing
  to an options, or ``specs'' file. \index{Snopt@\textsf{Snopt}!specs
  file} See the \SNOPT\ users guide to discover which options are
  available. If one chooses not to use an options file, \v{iSpecs}
  should be set to zero.  
  \index{iSpecs@{\texttt{iSpecs} (specs file descriptor)}!description}

  Default value $\v{iSpecs}= 0$.
  \index{iSpecs@{\texttt{iSpecs} (specs file descriptor)}!default value} 

 \parm{iPrint} is a \FORTRAN\ file descriptor pointing to a file that will
  be overwritten with the results of this run of \SNOPT. If one does not
  wish to save the output to a file, \v{iPrint} should be set to zero.
  \index{iPrint@{\texttt{iPrint} (print file descriptor)}!description}

  Default value $\v{iPrint}= 0$.
  \index{iPrint@{\texttt{iPrint} (print file descriptor)}!default value}

 \parm{iSumm} a \FORTRAN\ file descriptor pointing to a file that will
  be overwritten with summary information from this run of
  \SNOPT.  Typically, \v{iSumm} is either set to $6$, which will cause
  the summary output to be printed on the terminal, or set to $0$, which
  disables the printing of summary information.
  \index{iSumm@{\texttt{iSumm} (summary file descriptor)}!description}

  Default value $\v{iSumm}= 6$.
  \index{iSumm@{\texttt{iSumm} (summary file descriptor)}!default value}

 \parm{iErr} a \FORTRAN\ file descriptor pointing to a file that will be
  overwritten with diagnostic information from this run of \SNOPT. Set
  \v{iErr} to zero to disable printing of diagnostic information.
  \index{iErr@{\texttt{iErr} (diagnostic file descriptor)}!description}

  Default value $\v{iErr}= 0$.
  \index{iErr@{\texttt{iErr} (diagnostic file descriptor)}!default value} 

 \parm{cu, iu, ru} are character, integer, and double precision arrays in
  which the user may store information between calls to \usrfun.
  \index{cu@{\texttt{cu(lencu)} (user character scratch array)}!description}
  \index{iu@{\texttt{iu(leniu)} (user integer scratch array)}!description}
  \index{ru@{\texttt{ru(lenru)} (user double scratch array)}!description}
    
 \parm{cw, iw, rw} are character, integer, and double precision
  work-space arrays used by \SNOPT. These arrays must be declared
  sufficiently large for \SNOPT\ to solve the optimization problem.
  \index{cw@{\texttt{cw(lencw)} (\SNOPT\ character work array)}!description}
  \index{iw@{\texttt{iw(leniw)} (\SNOPT\ integer work array)}!description}
  \index{rw@{\texttt{rw(lenrw)} (\SNOPT\ double work array)}!description}
\end{parameters}

 \subsection{An Example Problem} \label{sec-userfuns-example}  

\index{usrini@{\texttt{usrini} (user-supplied subroutine)}!example}
\index{usrfun@{\texttt{usrfun} (user-supplied subroutine)}!example}
\index{prob@{\texttt{prob} files}!prob.f@\v{prob.f} (user supplied)}
\index{prob@{\texttt{prob} files}!fortran source@\FORTRAN\ source files} 

Here we give examples of subroutines \v{usrini} and \v{usrfun} for the following
four variable problem:
$$ \def\t{\phantom3}\def\o{\phantom1}\def\.{\phantom.}%
   \def\f{\phantom5}%
   \begin{array}{l@{\extracolsep{2ex}}r@{\extracolsep{0pt}}r%
        @{\extracolsep{0pt}}r%
        @{\extracolsep{4pt}}c@{\extracolsep{4pt}}r}
\minimize{}&{}    3x_1     &{}+ (x_1 + x_2 +  x_3)^2 &{}+    5x_4    &     &    \\
\subject   &             &        4x_2 + 2x_3    &             & \ge & 0\.\\
           &{}           &{}x_1 + x_2^2 + x_3^2    &             &  =  & 2\.\\
           &             &    x_2^4 + x_3^4      &{}+  \f x_4    &  =  & 4\.\\
           &             &x_1  \ge 0 \hss      &        x_4    & \ge & 0.
   \end{array}
$$
In the format of problem NP  we have $\Flow \le F(x) \le \Fupp$, where
$$
        \Flow = \mtx{l}{ -\infty  \\
                     \m  0      \\
                     \m  2      \\
                     \m  4        },
         \quad
        F = \mtx{c}{ 3x_1 + (x_1 + x_2 + x_3)^2 + 5x_4  \\
                     4x_2 + 2x_3                  \\
                     x_1  + x_2^2 + x_3^2           \\
                     x_2^4 + x_3^4    +   x_4         },
         \quad
        \Fupp = \mtx{l}{ +\infty  \\
                         +\infty  \\
                         \m  2   \\
                         \m  4     }.
\index{F@$F(x)$}\index{F1@$\Obj(x)$}\index{L@$\Flow$}\index{U@$\Fupp$}
$$
The objective function has been assigned to the first component of $F$,
which means that $\v{ObjRow} = 1$.  The objective component is not
constrained by \SNOPT, so there are infinite upper and lower bounds on
$\Obj$.  (A component with infinite upper and lower bounds is known as
a ``free row'' of the problem.)  \SNOPT\ automatically provides these
infinite bounds on the objective row, and so it is unnecessary to
provide them (unless later the user plans to set $\v{ObjRow}=0$ to make \SNOPT\
find a feasible point for the constraints).

The upper and lower bounds on the variables are given by $\xlow \le x 
\le\xupp$, where
$$
        \xlow = \mtx{l}{ \m   0  \\
                        -\infty  \\
                        -\infty  \\
                          \m   0   },
        \quad
        x = 
        \mtx{c}{ x_1 \\
                 x_2 \\
                 x_3 \\
                 x_4   },
        \quad
        \xupp = \mtx{l}{ +\infty  \\
                 +\infty  \\
                 +\infty  \\
                 +\infty    }.\index{l@$\xlow$} \index{u@$\xupp$} 
$$
Our version of subroutine \v{usrini} performs four tasks: (i) it defines the
length of the variable-dimensioned arrays used by \SNOPT\ and \SNADIOPT; (ii) it
opens the print file and summary file; (iii) it initializes the array of
variables; and (iv) it defines the upper and lower bounds on $x$ and $F$.
\begin{small}
\begin{verbatim}
      subroutine usrini( ObjAdd, ObjRow, Prob, 
     &     x, xlow, xupp, xstate, Names,
     &     Fmul, Flow, Fupp, Fstate, FNames,
     &     iSpecs, iPrint, iSumm, iErr,
     &     cu, iu, ru, cw, iw, rw )

      implicit           none
      integer            n, neF, nName, nFnames, ObjRow,
     &     lencw, leniw, lenrw, lencu, leniu, lenru
      parameter        ( lencw = 501, leniw = 10000, lenrw = 20000 )
      parameter        ( lencu =   1, leniu =     1, lenru =     1 )
      parameter        ( n     =   4, neF   =     4 )
      parameter        ( nName =   1, nFnames =   1 )
      integer            iSpecs, iPrint, iSumm, iErr, xstate(n),
     &     Fstate(neF), iu(leniu), iw(leniw)
      double precision   ObjAdd, x(n), xlow(n), xupp(n), Flow(neF),
     &     Fupp(neF), Fmul(neF), ru(lenru), rw(lenrw)
      character*8        Prob, Names(nName), FNames(nFnames),
     &      cu(lencu), cw(lencw)
*     ==================================================================
*     usrini defines input data for the problem discussed in the
*     SnadiOpt Users Guide.
*     ==================================================================
      integer            i
      character*20       lfile
      double precision   plInfy
      parameter         (plInfy = 1.0d+20 )
*     ------------------------------------------------------------------
*     Initial x.

      x(1)   =  1.0d+0
      x(2)   =  1.0d+0 
      x(3)   =  1.0d+0
      x(4)   =  1.0d+0

      xlow(1) = 0.0d+0
      xlow(4) = 0.0d+0

*     Impose bounds on the constraint rows.

      Flow(2)   =  0.0d+0
      Flow(3)   =  2.0d+0       ! Equality constraint
      Fupp(3)   =  2.0d+0
      Flow(4)   =  4.0d+0       ! Equality constraint
      Fupp(4)   =  4.0d+0

      iSpecs = 4
      iPrint = 15
     
      lfile = 'prob.spc'
      open( iSpecs, file=lfile, status='OLD',     err=800 )
      lfile = 'prob.out'
      open( iPrint, file=lfile, status='UNKNOWN', err=800 )

      return
      
  800 write(iErr, 4000) 'Error while opening file', lfile
 4000 format(/  a, 2x, a  )

      end ! subroutine usrini
\end{verbatim}
\end{small}
Note that default initial values are used for the variables \v{Prob}, \v{Fmul},
\v{xstate}, \v{Fstate}, and \v{ObjAdd}.  Similarly, only those bounds not equal
to their default infinite values are assigned.

The subroutine \v{usrfun} defines the values of the vector $F(x)$.
\begin{small}
\begin{verbatim}
      subroutine usrfun( Status, mode, 
     &     neF, n, x, F,
     &     cu, lencu, iu, leniu, ru, lenru, 
     &     cw, lencw, iw, leniw, rw, lenrw )

      implicit           none
      integer            Status, mode, neF, n, lencu, leniu, lenru,
     &     lencw, leniw, lenrw, iu(leniu), iw(leniw)
      double precision   F(neF), x(n), ru(lenru), rw(lenrw)
      character*8        cu(lencu), cw(lencw)
*     ==================================================================
*     Usrfun computes the objective and constraint functions for the
*     problem featured in the SnadiOpt users guide.
*     ==================================================================
      integer            Obj
*     ------------------------------------------------------------------
      Obj  = 1                  ! The objective row

      F(Obj) = 3.0d+0*x(1) + (x(1) + x(2) + x(3))**2 + 5.0d+0*x(4)

*     Constraint functions.

      F(2)   =              4.0d+0*x(2)    + 2.0d+0*x(3)
      F(3)   =       x(1) +        x(2)**2 +        x(3)**2 
      F(4)   =                     x(2)**4 +        x(3)**4  + x(4)

      end ! subroutine usrfun 
\end{verbatim}
\end{small}

\index{function definition routine|see{\v{usrfun}}}
\index{user-supplied subroutines|see{\v{usrfun},  \v{usrini}}}
\index{Lagrange multipliers|see{\v{Fmul}}}
\index{shadow prices|see{\v{Fmul}}}
\index{dual variables|see{\v{Fmul}}}
\index{variable names|see{\v{Names}}} 
\index{constraint names|see{\v{Fnames}}} 
\index{variables|see{\v{x}}}
\index{data initialization routine|see{\v{usrini}}}
\index{summary file|see{\v{iSumm}}}
\index{bounds on \v{F}|see{\v{Flow} and \v{Fupp}}}
\index{bounds on \v{x}|see{\v{xlow} and \v{xupp}}}
\index{dual variables|see{\v{Fmul}}}
\index{specs file|see{\SNOPT\ specs file}}
\index{vector of variables|see{\v{x}}}
\index{vector of constraints|see{\v{F}}}
\index{options file|see{\SNOPT\ specs file}}
\index{Snopt@\textsf{Snopt}!specs file|see{{\itshape also} \v{iSpecs}}}
\index{Snopt@\textsf{Snopt}!work-space|see{work arrays}}
\index{Snopt@\textsf{Snopt}!user work-space|see{scratch arrays}}


\section{Invoking SnadiOpt} \label{invoking}

The user-supplied routines must be run through \ADIFOR\ and then
compiled and linked into a complete program, before \SNOPT\ may be
invoked. There are two steps in the process of building this complete
program. First, the user invokes the script \v{snadiopt.pl} to scan
\index{snadiopt.pl@\texttt{snadiopt.pl}!invocation}
the user-supplied routines and produce the components that are
needed to build a complete executable, notably the main program
itself. Second, the user invokes a version of the program \v{make}
to perform the build.
\index{Gmake@{\textsf{GNU Make}}}
\index{Make@\textsf{Make}!used to build executables} 

Users will typically call \v{snadiopt.pl} only once. Most changes
made to a model can be incorporated into the executable by simply
typing \v{make}. By design, \v{snadiopt.pl} generates a
(relatively) straightforward set of components that may be modified at
will. In some cases, particularly if the sizes of the arrays defined in
\v{usrini} change, \index{usrini@{\texttt{usrini} (user-supplied subroutine)}!changing the source}
it may be convenient to call \v{snadiopt.pl}
again rather than modifying multiple files.  We have incorporated a
feature into \v{snadiopt.pl} to simplify this process. See
Section~\ref{merging} for more information.

\subsection{Locating Executables and Libraries}  
\label{locating}

Before one can use \SNADIOPT, the package must, of course, be
installed on the user's machine. Installation
instructions are provided with the \SNADIOPT\ distribution. Because
the installation process depends on the machine type and the source of
the distribution, installation instructions will not be repeated
here. A few words, however, are in order.

The \SNADIOPT\ package consists of a program named \v{snadiopt.pl},
some data files needed by \v{snadiopt.pl}, and code libraries
that must be combined (linked) with user-supplied code to produce a
problem executable. Ideally, these components will be installed in an
appropriate, system-dependent location. On \UNIX\  systems, for instance,
the default location is the \v{/usr/local/} directory structure. If
the \SNADIOPT\ package is located in some appropriate system location,
then typing  \index{UNIX@\textsf{UNIX}} \index{snadiopt.pl@\texttt{snadiopt.pl}!help option}
\begin{verbatim}
% snadiopt.pl --help
\end{verbatim}
will provide summary help information about using the \v{snadiopt.pl}
program.  If this is the case, then the rest of this section may be 
skipped. If the system cannot find the \v{snadiopt.pl} program, then
the user will need to tell the system where to find it. The following
instructions are for versions of the \UNIX\
operating system.

First, one should ask the individual who installed \SNADIOPT\ where 
the \v{snadiopt.pl} program is located.  If it is not in a 
system-dependent location, it will normally be found in the \v{bin} 
subdirectory of the \SNOPT\ distribution.  The \v{cd} command may be 
used to change the working directory to the directory containing 
\v{snadiopt.pl}.  The command used to set the  \v{PATH} environment
variable depends on which shell is being used.
For \v{bash} or \v{ksh}, the appropriate command is
\begin{verbatim}
% PATH=$PATH:$PWD
\end{verbatim}
but for \v{csh} or \v{tcsh}, the command 
\begin{verbatim}
% setenv PATH ${PATH}:${PWD}
\end{verbatim}
must be used. Virtually all users will
be using a shell that responds to one of these two commands. It is
safe to try these commands if one is unsure which shell is being used.

Once the \v{PATH} environment variable has been set, the system will 
be able to find \v{snadiopt.pl}. One can
then generate and compile problem executables. The executables may
not, however, run. While this would seem to be undesirable behavior,
there is actually a reason for it. Modern operating systems support
the concept of dynamically linked libraries. Such libraries are not
copied into an executable, but rather are loaded into memory when a
program is run. With such a scheme, several executables may share one
library. \SNADIOPT\ uses dynamically linked libraries whenever
possible.

Because dynamically linked libraries must be loaded at run-time, the 
operating system must know where to locate these libraries.  The 
simplest scheme is to place the libraries in a system-dependent 
location.  If one had to set the \v{PATH} environment variable to 
tell the system where to find the \v{snadiopt.pl} program, then it is 
likely that one will need to tell the system where to find the 
\SNADIOPT\ libraries as well.  To do so, one should first change the 
working directory to the directory containing the \SNADIOPT\ 
libraries.  These libraries will usually be located in the \v{lib} 
subdirectory of the \SNOPT\ distribution, and will have names similar 
to \v{libsnddiopt.so} and \v{libsnsdiopt.so}.  If the user is using 
\v{bash} or \v{ksh},  the command
\begin{verbatim}
% LD_LIBRARY_PATH=$LD_LIBRARY_PATH:$PWD
% export LD_LIBRARY_PATH
\end{verbatim}
should be executed. Those using \v{csh} or
\v{tcsh}, must determine whether \v{LD\_LIBRARY\_PATH} has
already been set. If the command
\begin{verbatim}
% printenv LD_LIBRARY_PATH
\end{verbatim}
prints nothing, then the variable has not been set, and 
\begin{verbatim}
% setenv LD_LIBRARY_PATH ${PWD}
\end{verbatim}
will set it appropriately. Otherwise, typing
\begin{verbatim}
% setenv LD_LIBRARY_PATH ${LD_LIBRARY_PATH}:${PWD}
\end{verbatim}
will add the current directory to the existing \v{LD\_LIBRARY\_PATH}.

Let us make a few, final notes about this process.  These days, it is
common for a user to have multiple command windows open. It is a
frequent mistake to think that setting the \v{PATH} variable, or
any variable, in one window sets its value in all windows. Setting a
variable in one command window does not affect the other command
windows in any way.  It is usually possible to alter certain
initialization files to set the values of \v{PATH} and
\v{LD\_LIBRARY\_PATH} in every command window automatically at login.
It is not possible for us to describe this process, because it is very
system dependent.  One should ask a system administrator how to do this.

\subsection{Basic Usage}
\label{snadiopt-basic-usage}

Suppose a user has placed all the code needed to define a certain
problem, including the required subroutines \v{usrfun} and 
\v{usrini}, in a file named \v{prob.f}. The command
\begin{verbatim}
% snadiopt.pl -o prob prob.f
\end{verbatim}
\index{prob@{\texttt{prob} files}!prob (executable)@\v{prob} (executable)}
\index{prob@{\texttt{prob} files}!prob.f@\v{prob.f} (user supplied)} will
generate the files that are needed to build an executable named
\v{prob} that solves the user's optimization model.

To actually build the executable, invoke the \GNU\ version of the
program \v{make}. \index{Gmake@{\textsf{GNU Make}}} On many systems,
\GNU\ make is installed as \v{make} or \v{gmake}, and so typing
\begin{verbatim}
% make
\end{verbatim}
or
\begin{verbatim}
% gmake
\end{verbatim}
should build a program named \v{prob} that may be executed from the
command line.

 \subsection{Files Generated by \v{snadiopt.pl}} \label{generated-files}
 \index{ADIFOR@\textsf{ADIFOR}!generated files|(}

In this section, we briefly describe the files generated by
\v{snadiopt.pl} itself. \index{snadiopt.pl@\texttt{snadiopt.pl}!generated files}
Other temporary files may be generated by
the compiler and \ADIFOR.\ A beginning user should not need to know
about the generated files in order to use this package. Therefore,
this section may be skipped on a first reading.

This section lists the files generated for a problem named
\v{prob}. 
\index{prob@{\texttt{prob} files}!prob (executable)@\v{prob} (executable)}
The script \v{snadiopt.pl} uses ``prob'' as the
prefix for most of the files generated for this problem. If the
user had invoked the command
\begin{verbatim}
% snadiopt.pl -o prob prob.f
\end{verbatim}
then ``prob'' will be the prefix of the generated files. In general,
the \v{-o} option determines the prefix of the generated files. If the
option is omitted, then ``unnamed'' will be the prefix used.

\paragraph{\GNUmakefile.} \index{makefile@\texttt{makefile}!\v{GNUmakefile}}
This is the only generated file that is not prefixed by the name of
the problem. The \GNUmakefile\ is meant to be shared by all problems
in a given directory; it contains general information about building
and managing problem executables.

Users may wish to modify \GNUmakefile\ to customize the build
process. For instance, \GNUmakefile\ might be modified to tell the
compiler to generate object code suitable for use with a debugger. The
\v{snadiopt.pl} program will not overwrite an existing \GNUmakefile,
unless it is called with the \v{--refresh-makefile} option. Therefore,
it is safe to modify this file.

\paragraph{\v{prob\_submake}, \v{prob\_submake.orig}, \v{prob\_submake.bak}.} 
\index{prob@{\texttt{prob} files}!prob\_submake@\v{prob\_submake}}
The file \v{prob\_submake}
contains the commands for building the program, including the commands
for calling \ADIFOR\ \index{ADIFOR@\textsf{ADIFOR}!generated files}.
The file should contain the complete dependency information for the
program and should be capable of rebuilding the program when components are
modified.

It is sometimes necessary to modify \v{prob\_submake}. The file
\v{prob\_submake.orig}  contains the original version of this file, as
generated by \v{snadiopt.pl}.  This allows the user to compare the
modified version of \v{prob\_submake} with the original file.

If \v{snadiopt.pl} detects an existing file named \v{prob\_submake},
it will save this file as \v{prob\_submake.bak}. The user may then reapply
any changes made to \v{prob\_submake} to the newly generated file. Many
times, these changes can be merged automatically.  See
Section~\ref{merging} for more information.

\paragraph{\v{prob\_main.f}, \v{prob\_main.f.orig}, \v{prob\_main.f.bak}.}
\index{prob@{\texttt{prob} files}!prob\_main@\v{prob\_main}}
\index{prob@{\texttt{prob} files}!prob\_main.f@\v{prob\_main.f.orig}}
\index{prob@{\texttt{prob} files}!prob\_main.f.bak@\v{prob\_main.f.bak}}
The file \v{prob\_main.f} contains the \FORTRAN\ main program that
calls \SNOPT\ with the user's data and problem definition
functions. It also performs some necessary bookkeeping and
initialization and is responsible for allocating the arrays that the
user requests in the \v{usrini} subroutine.

Users may wish to modify \v{prob\_main.f}. For instance, a user might
wish to output the results from \SNOPT\ in a particular format and so
might place the commands for doing so in \v{prob\_main.f}. The file
\v{prob\_main.f.orig} contains the version of \v{prob\_main.f}
generated by the last call to \v{snadiopt.pl}. This allows the user to
compare the modified \v{prob\_main.f} with the original file.

If \v{snadiopt.pl} detects an existing file named \v{prob\_main.f},
it will save that file as \v{prob\_main.f.bak} before proceeding.  All
changes that the user had made to the existing \v{prob\_main.f} are
saved in that back-up file, and the user may reapply these changes to
the newly generated file.  Many times, these changes can be merged
automatically.  See Section~\ref{merging} for more information.

\paragraph{\v{prob.adf}.}
\index{prob@{\texttt{prob} files}!prob.adf@\v{prob.adf}} The file \v{prob.adf} 
contains the \ADIFOR\ \index{ADIFOR@\textsf{ADIFOR}!generated files} 
``script'' for differentiating the model's functions.  See the 
\ADIFOR\ users guide for more information.  It is unlikely that a 
user will need to modify this file.

\paragraph{\v{prob\_admain.f}.}
\index{prob@{\texttt{prob} files}!prob\_admain.f@\v{prob\_admain.f}}
\ADIFOR\ requires a complete compilable program in order to
differentiate a function called from that program. The file
\v{prob\_admain.f} contains a phony program that calls \v{usrfun}. We
know of no reason for users to modify this file.
\index{usrfun@{\texttt{usrfun} (user-supplied subroutine)}!called by\v{prob\_admain.f}}

\paragraph{\v{prob\_sparse\_dispatch.f}, \v{prob\_dense\_dispatch.f}.}
\index{prob@{\texttt{prob} files}!prob\_sparse@\v{prob\_sparse\_dispatch.f}}
\index{prob@{\texttt{prob} files}!prob\_dense@\v{prob\_dense\_dispatch.f}} 
These files call library routines supporting the use of \ADIFOR\ with
\SNOPT. The existence of these files is an artifact of the \FORTRAN\ 
language not having syntax for storing a reference to a subroutine. We
know of no reason for users to modify these files.

\paragraph{\v{prob.cmp}.}
\index{prob@{\texttt{prob} files}!prob.cmp@\v{prob.cmp}}
 The file \v{user.cmp} 
is not generated by \v{snadiopt.pl}, but rather is created as an 
intermediate file in the build process.  It contains a list of 
\FORTRAN\ files that are to be sent to \ADIFOR. One should not not 
modify this file; one should modify the \v{AD\_SOURCE} and 
\v{AD\_OTHER\_FILES} variables in the file \v{prob\_submake}.  
\index{AD\_SOURCE@\texttt{AD\_SOURCE}} 
\index{AD\_OTHER\_FILES@\texttt{AD\_OTHER\_FILES}}

 \subsection{Merging Changes} \label{merging} 

Users need to call \v{snadiopt.pl} only once. The components that it
creates may then be modified at will, and the executable rebuilt using
\v{make}. However, on some occasions it may be useful
to call \v{snadiopt.pl} again,
\index{snadiopt.pl@\texttt{snadiopt.pl}!merging changes} particularly
when
\begin{itemize}
 \item the sizes of the arrays in \v{usrini} have changed.
 \index{usrini@{\texttt{usrini} (user-supplied subroutine)}!merging changes}
  The array sizes in \v{prob\_main.f} 
  \index{prob@{\texttt{prob} files}!prob\_main.f@\v{prob\_main}}
 and possibly \v{prob.adf} 
  \index{prob@{\texttt{prob} files}!prob.adf@\v{prob.adf}}
  must also be
  modified. The program \v{snadiopt.pl} will update these
  quantities automatically.
  
\item the names of the parameters of \v{usrfun} have changed.
  \index{usrfun@{\texttt{usrfun} (user-supplied subroutine)}!merging
    changes} The user must either call \v{snadiopt.pl} again, or edit
  \v{prob.adf} to update the names of the independent variables
  (\v{AD\_IVARS}) and the names of the dependent variables
  (\v{AD\_DVARS}).

\item the names of \FORTRAN\ source files are modified, or new source
  files are added. Users will need to update \v{prob\_submake},
  \index{prob@{\texttt{prob} files}!prob\_submake@\v{prob\_submake}} 
or call \v{snadiopt.pl}
  again.
\end{itemize}

\index{prob@{\texttt{prob} files}!prob\_submake@\v{prob\_submake}} 
\index{prob@{\texttt{prob} files}!prob\_main@\v{prob\_main}}

It is not uncommon, however, for users to want to modify
\v{prob\_submake} to customize the build process, or modify
\v{prob\_main.f} to perform some action on the results of
\snopt. Normally, when \v{snadiopt.pl} is called, it overwrites these
files, saving copies of the existing files as \v{prob\_main.f.bak} and
\v{prob\_submake.bak}. Users may then reapply the changes they had
made to the old \v{prob\_main.f} and \v{prob\_submake} to the newly
generated files.

There are, however, \UNIX\ utilities \index{UNIX@\textsf{UNIX}!merge
option} that are able to merge changes between versions automatically.
The \v{snadiopt.pl} \v{--merge} option provides an interface to these
tools.  Simply call \v{snadiopt.pl} with the arguments
\begin{verbatim}
% snadiopt.pl --merge -o prob prob.f
\end{verbatim}
\index{snadiopt.pl@\texttt{snadiopt.pl}!\v{merge option}} The merge is
based only on the comparison of blocks of text. It does not pretend to
understand the meaning of the code. However, it is effective
remarkably often. In case the merge is ineffective, the files that
\v{snadiopt.pl} would have produced without the merge option may be
found in \v{prob\_submake.orig} and \v{prob\_main.f.orig}.

Rarely, there will be conflicts that make it impossible to
complete the merge. In these cases, lines of the form
\begin{verbatim}
<<<<<<< prob_submake.bak
lines from prob_submake.bak
=======
lines from prob_submake.orig
>>>>>>> prob_submake.orig
\end{verbatim}
will be inserted in the files, and these sections must be edited by hand.

The merge option uses the standard \UNIX\ utilities \v{diff3} and
\v{ed}.  \index{UNIX@\textsf{UNIX}!\v{diff3} and \v{ed}}
\index{UNIX@\textsf{UNIX}!merge option} Merging is not supported on
platforms on which these programs are not available. We don't support
automatic merging of the other generated files. Merging requires that
the files \v{prob\_main.f.orig} and \v{prob\_submake.orig} generated
by the last call to \v{snadiopt.pl} be present in the current
directory. \index{snadiopt.pl@\texttt{snadiopt.pl}!merging changes}

 \index{ADIFOR@\textsf{ADIFOR}!generated files|)}

\subsection{Advanced Usage}
\index{source files|see{\v{prob} files}}

\paragraph{Multiple Source Files.}
\index{snadiopt.pl@\texttt{snadiopt.pl}!loading multiple source files}
\index{prob@{\texttt{prob} files}!combining multiple source files}
The script \v{snadiopt.pl} is not restricted to scanning a single
file. If several \FORTRAN\ files are needed to define the problem,
all file names should be included on the command line.

\paragraph{Library Source Files.}
\index{ADIFOR@\textsf{ADIFOR}!library source files}
\index{Fortran@{\textsf{Fortran} compiler}!intrinsics}
\ADIFOR\ understands \FORTRAN\ intrinsics, operators such as \v{sqrt}
that look like functions but actually have special status in the
language. It must, however, have the source code to actual functions
used in the program, such as the functions defined in the
\BLAS~\cite{blas1979}. Source files for these
functions must be included on the command line. 
\index{BLAS@{Basic Linear Algebra Subroutines, \textsf{BLAS}}}

If a user is certain that the included library functions do not need
to be differentiated, and would rather link against the installed
library than recompile, he may include the source file names in the
\v{AD\_OTHER\_FILES}
\index{AD\_OTHER\_FILES@\texttt{AD\_OTHER\_FILES}} variable in the
\v{prob\_submake}.  \index{prob@{\texttt{prob}
    files}!prob\_submake@\v{prob\_submake}} See Section~\ref{makefile}
for more information.

\paragraph{Using Alternative Function Names} The \v{snadiopt.pl}
script tries to be flexible about the names of the problem definition 
functions.  Several options that allow these names and the names of 
certain output files to be specified by the user.  The available 
options are summarized in Section~\ref{summary-of-options}.\index{usrini@{\texttt{usrini} (user-supplied 
subroutine)}!using an alternate name} \index{usrfun@{\texttt{usrfun} 
(user-supplied subroutine)}!using an alternate name}

\index{snadiopt.pl@\texttt{snadiopt.pl}!options}
\subsection{Summary of All Options}
\label{summary-of-options}
\begin{verbatim}
Usage: snadiopt.pl [switches] file1.f [file2.f]
   -help                 Print this message.
   -version              Print the version number of snadiopt.pl.
   -o PROGRAM            The optimization problem (and binary executable) 
                         will be named PROGRAM. (default: a.out)
   -makefile MAKEFILE    The output makefile will be named MAKEFILE.
                         (default: PROGRAM_submake or unnamed_submake
                         if PROGRAM is not specified.)
   -refresh-makefile     Create MAKEFILE, even if it already exists.
                         Unless given this option, the script will not
                         overwrite an existing MAKEFILE.
   -usrfun NAME          The FORTRAN subroutine named NAME computes the
                         functions needed in this optimization problem.
                         (default: usrfun)
   -usrini NAME          The FORTRAN subroutine named NAME initializes
                         this optimization problem. (default: usrini)
   -merge                Merge changes between prob_main.f.orig and 
                         the current version of prob_main.f into the newly
                         generated prob_main.f. Do the same for
                         prob_submake.
\end{verbatim}


 \section{Building the Executable} \label{makefile}

The \UNIX\ \MAKE\ utility is used to generate targets, 
in this case executables that solve specific optimization problems,
from source files. The rules that \MAKE\ uses to build these targets
are specified in files known as makefiles.  The \MAKE\  utility is
also commonly used to perform certain bookkeeping tasks, such as
removing files generated by the build process.
\index{Make@\textsf{Make}!\GNU\ version of \v{make}}
\index{Gmake@{\textsf{GNU Make}}}

This project uses the \GNU\ dialect of \MAKE.  This dialect has
certain pattern substitution features that are absent in other
versions of \MAKE. Furthermore, \GNU\ \MAKE\ is freely available on
virtually every platform. Vendor-specific versions of \MAKE\ are not
consistent in interface, language, or quality. Thus, we use \GNU\ make
to get predictable performance on a wide variety of platforms.

We assume that the reader has a basic knowledge of the \MAKE\
utility. (For a good introduction to \MAKE, see
\cite{managing_projects_make} or \cite{gnu_make}.) This section
describes how we have arranged our makefiles, the targets that are
available, and variables that may be modified to effect the build
process. Most users will simply invoke \GNU\ \MAKE\ without any
arguments to build all problems in the current directory (provided
that \v{snadiopt.pl} has already been invoked to create the necessary
components.)  \index{snadiopt.pl@\texttt{snadiopt.pl}!build process}
For the rest of this section, we assume that \GNU\ \MAKE\ has
been installed and may be invoked by the command \v{make}.  Users 
should substitute
the command that they use to invoke \GNU\ \MAKE\ wherever appropriate.

On \UNIX\  systems, if the \SNADIOPT\ package is not
installed in an appropriate system location, the user may have to set
the \v{LD\_LIBRARY\_PATH} environment variable before the executables
that are built will run. See Section~\ref{locating} for instructions on
how to do this.

 \subsection{Typical Usage}

Before invoking \MAKE, users must call \v{snadiopt.pl}
\index{snadiopt.pl@\texttt{snadiopt.pl}!build process} to generate the components of
each problem they wish to build. Then, typing \v{make} will cause
executables to be built for all models in the current directory. If
users  wish to build executables for only some of the problems, they
may instead list the names of the executables that they wish to
build. For instance;
\begin{verbatim}
% make prob1 prob3
\end{verbatim}
would build only the executables \v{prob1} and \v{prob3}.

\subsection{Subordinate Makefiles}
\index{makefile@\texttt{makefile}!subordinate}

Traditionally, \MAKE\ takes all its input from a single file,
typically named \v{makefile}, \v{Makefile}, or \v{GNUmakefile}.  This
scheme has proven to be too restrictive in practice, so many versions
of \MAKE, and \GNU\ make in particular, support the \v{include}
directive.
\index{makefile@\texttt{makefile}!include directive}
A line of the form  \index{makefile@\texttt{makefile}!include directive}
\begin{verbatim}
include filename
\end{verbatim}
tells \MAKE\ to act is if all the text in ``filename'' were included literally
in the makefile.

In a problem directory, there will be a single file named \GNUmakefile\
and one or more files with the suffix ``submake.'' 
\index{makefile@\texttt{makefile}!use of submake}
Each of the files with an appended \v{\_submake} is called a subordinate
makefile, because it does not contain a complete set of rules
and dependencies for building the executable for a problem. The
\GNUmakefile\ uses the \v{include} directive to include the text of
all the subordinate makefiles. Each model in a directory will have its
own subordinate makefile, which will contain the specific rules,
variables, and dependencies for building an executable that optimizes
that model. The text of \GNUmakefile\ contains generic rules and
dependencies that are needed to build any model.

Subordinate makefiles are useful for several reasons:
\begin{itemize}
 \item Users may wish to have more than one model in a directory.
  Having complete, separately named makefiles for each model becomes
  awkward, requiring the user to specify the name of the makefile for
  each build.

 \item Users may want to build the executable for more than one model
  or to take some other action that affects more than one model. When
  subordinate makefiles are used, the rules for all the models are
  available, so a user may build any combination of targets by typing
  their names on the command line. The command ``\v{make all}'' works
  as expected and is the default target.
  \index{makefile@\texttt{makefile}!targets!all@\texttt{all}}
  
 \item Sometimes multiple models will share one or more files. Because
  \MAKE\ is given the complete set of dependencies for the executables
  of all the models, it can quickly determine which files need to be
  rebuilt.  If each model had a separate makefile, the user would have
  to make this determination or, alternatively, call \MAKE\ once for
  every executable.

\end{itemize}

The \MAKE\ program can scan all the subordinate makefiles and build a
complete set of dependencies quickly. The time taken is typically
many times shorter than the time needed to compile a single file. Some
users may be surprised by this speed.  See~\cite{recursive_make} for a
discussion of issues affecting the efficiency of \MAKE.
\index{makefile@\texttt{makefile}!recursive invocation}

\subsection{Useful \texttt{makefile} Targets}

In addition to the names of the programs, a number of
``phony'' targets may be specified for \MAKE. These
targets cause some action to be taken, rather than causing the target
to be built. These targets commonly are defined to perform
useful project-management tasks, such as deleting ``\v{.o}'' files.
  \index{makefile@\texttt{makefile}!targets!phony@\texttt{phony}}

\begin{description}

\item[\v{all}] builds everything. This is the default, so \v{make all}
  is equivalent to \MAKE.
  \index{all@\texttt{all}|see{make targets}}
  \index{makefile@\texttt{makefile}!targets!all@\texttt{all}}

  \item[\v{check}] checks the consistency of the \FORTRAN\ files used 
  to build each program.  This requires that \v{ftnchek} 
  \index{ftnchek@\textsf{ftnchek}} has been installed.  The 
  \v{ftnchek} program and documentation are freely available and may 
  be obtained from \v{http://www.netlib.org}.  
  \index{ftnchek@\textsf{ftnchek}!Web page} 
  \index{makefile@\texttt{makefile}!targets!check@\texttt{check}}

\item[\v{clean}] removes object (``\v{.o}'') files and some common "garbage"
  files, such as \v{core} files. This does not remove the executable file or
  any files generated by \ADIFOR.
  \index{makefile@\texttt{makefile}!targets!clean@\texttt{clean}}
  \index{ADIFOR@\textsf{ADIFOR}!removing \v{*.o} files}

\item[\v{veryclean}] removes more files generated by the build process,
  including the executable and all output from \ADIFOR. This target
  also removes files named ``\v{prob.out}'', the traditional name for
  output from the solvers.
  \index{makefile@\texttt{makefile}!targets!veryclean@\texttt{veryclean}}
  \index{prob@{\texttt{prob} files}!prob.out@\v{prob.out}}
 
\item[\v{distclean}] cleans up for distribution.  This target is like
  \v{veryclean} but does not delete the differentiated \FORTRAN\ 
  problem files, since those files are considered part of a distribution. Use
  this target to distribute the files to a machine on which
  the \ADIFOR\ translator is not available. This target invokes
  \v{clean}, \v{adifor-clean} and \v{snadiopt-clean}.
  \index{makefile@\texttt{makefile}!targets!distclean@\texttt{distclean}}

\item[\v{maintainer-clean}] deletes everything that can be rebuilt. This
  includes the files created by a call to \v{snadiopt.pl} and the
  makefiles themselves.
  \index{makefile@\texttt{makefile}!targets!maintainer-clean@\texttt{maintainer-clean}}

\item[\v{adifor-clean}]  removes \ADIFOR\ auxiliary files, but not the
  autodifferentiated \FORTRAN\ files.
  \index{makefile@\texttt{makefile}!targets!adifor-clean@\texttt{adifor-clean}}
  \index{ADIFOR@\textsf{ADIFOR}!removing \ADIFOR\ auxiliary files}

\item[\v{adifor-veryclean}] removes \ADIFOR\ auxiliary files and 
  the differentiated \FORTRAN\ files  (and \v{*.cmp}).
  \index{makefile@\texttt{makefile}!targets!adifor-veryclean@\texttt{adifor-veryclean}}
  \index{ADIFOR@\textsf{ADIFOR}!removing \ADIFOR\  differentiated files}
   
\item[\v{snadiopt-clean}] removes auxiliary files generated by
  \SNADIOPT\ for use with all the programs. These are the
  ``\v{.bak}'' and ``\v{.orig}'' files.
  \index{makefile@\texttt{makefile}!targets!snadiopt-clean@\texttt{snadiopt-clean}}

 \item[\v{snadiopt-veryclean}] removes all files generated by \SNADIOPT\ 
   for use with all the programs.  Files generated by \SNADIOPT\ 
   cannot be rebuilt by using commands in the makefile. This target is
   intended to reverse the effect of calling \v{snadiopt.pl}.
  \index{makefile@\texttt{makefile}!targets!snadiopt-veryclean@\texttt{snadiopt-veryclean}}
  \index{SnadiOpt@{\textsf{SnadiOpt} package}!make targets}

\end{description}

These targets also have versions that are 
limited to the components of a single module. For instance, 
\v{make prob-clean}
\index{makefile@\texttt{makefile}!targets!prob-clean@\texttt{prob-clean}}
will remove auxiliary files generated in a build of the executable
\v{prob}. In general, any of these targets may be prefixed by the
name of a specific problem. 
\index{prob@{\texttt{prob} files}!prob (executable)@\v{prob} (executable)}

 \subsection{Useful \texttt{makefile} Variables} 

Each subordinate makefile contains the following variables that users
might need to modify.

\begin{description}
 \item[\v{prob\_USER\_LIBS}] defines any libraries that need to be
 linked with the user's code to produce an executable. The \SNADIOPT\
 libraries are automatically included.
 \index{prob\_USER\_LIBS@\texttt{prob\_USER\_LIBS}} 
 \index{libraries!\ADIFOR}\index{libraries!\SNADIOPT}
 \index{ADIFOR@\textsf{ADIFOR}!libraries}\index{SnadiOpt@{\textsf{SnadiOpt} package}!libraries}
  
 \item[\v{prob\_SOURCE}] is a list of all \FORTRAN\ source files for
the model \v{prob}. A reference to the variables \v{prob\_AD\_SOURCE}
and \v{prob\_AD\_G\_SOURCE} should appear in this list.
\index{prob\_SOURCE@\texttt{prob\_SOURCE}} 
\index{prob@{\texttt{prob} files}!fortran source@\FORTRAN\ source files} 

 \item[\v{prob\_AD\_SOURCE}] is the list of files to be differentiated by
  \ADIFOR. 
  \index{prob\_AD\_SOURCE@\texttt{prob\_AD\_SOURCE}}
  \index{ADIFOR@\textsf{ADIFOR}!files for model \v{prob}} 

 \item[\v{prob\_AD\_OTHER\_FILES}] is a list of files that must be
passed to \ADIFOR\ in addition to those of \v{prob\_AD\_SOURCE} in
order to make a complete program. The files in this list differ from
the files in \v{prob\_AD\_SOURCE} in that neither the original file
nor the result computed by \ADIFOR\ need be compiled into the problem
executable. Phony main programs and phony library stubs belong on this
list.
  \index{prob\_AD\_OTHER\_FILES@\texttt{prob\_AD\_OTHER\_FILES}} 
  \index{ADIFOR@\textsf{ADIFOR}!linking other files}

\end{description}

\subsection{Dense \ADIFOR}
\index{dense \ADIFOR} 
\index{ADIFOR@\textsf{ADIFOR}!for dense Jacobians} 

\SNOPT\ is a  sparse optimization solver;%
\index{Snopt@\textsf{Snopt}!sparse optimization}%
\index{nonlinear constrained optimization}%
\index{nonlinear programming}
its internal data is
stored in sparse matrix format.  Sparse matrix format is designed to
take advantage of the fact that many the elements of the matrix will
be identically zero. \index{Jacobian matrix (J)!sparse matrix format}

 \ADIFOR\ can either generate derivative code that uses sparse
matrices internally or code that uses dense matrices internally. For
small problems, typically problems with fewer than 30 variables, the
derivative code generated by dense \ADIFOR\ can be more
efficient. This is typically not an important issue unless the problem
is highly nonlinear (otherwise, it is just a small simple problem
and will be solved quickly regardless of which version of \ADIFOR\ is
used.) 
\index{ADIFOR@\textsf{ADIFOR}!sparse vs. dense} 

A sequence of commands of the form
\begin{verbatim}
% make adifor-veryclean
% make AD_FLAVOR=dense
\end{verbatim}
\index{makefile@\texttt{makefile}!targets!adifor-clean@\texttt{adifor-clean}}
\index{AD\_FLAVOR@\texttt{AD\_FLAVOR}}
will cause \ADIFOR\ to generate code that uses dense matrices 
\index{dense matrix format} to compute derivatives. It is important to 
\v{make adifor-veryclean}  \index{makefile@\texttt{makefile}!targets!adifor-veryclean@\texttt{adifor-veryclean}}
whenever switching between dense and sparse versions of \ADIFOR. 
The \v{make} program is unable to tell that a user has switched 
versions of \ADIFOR, \index{ADIFOR@\textsf{ADIFOR}!sparse vs. dense}
and it thus cannot tell which files need to be rebuilt.

\index{Makefile@\texttt{Makefile}|see{\v{makefile}}}
\index{GNUmakefile@\texttt{GNUmakefile}|see{\v{makefile}}}
\index{gmake@\texttt{gmake}|see{\GNU\ \MAKE}}
\index{subordinate makefile|see{\v{makefile}}}
\index{check@\texttt{check}|see{\v{makefile} targets}}
\index{clean@\texttt{clean}|see{\v{makefile} targets}}
\index{veryclean@\texttt{veryclean}|see{\v{makefile} targets}}
\index{all@\texttt{all}|see{\v{makefile} targets}}
\index{distclean@\texttt{distclean}|see{\v{makefile} targets}}
\index{maintainer-clean@\texttt{maintainer-clean}|see{\v{makefile} targets}}
\index{adifor-clean@\texttt{adifor-clean}|see{\v{makefile} targets}}
\index{adifor-veryclean@\texttt{adifor-veryclean}|see{\v{makefile} targets}}
\index{snadiopt-clean@\texttt{snadiopt-clean}|see{\v{makefile} targets}}
\index{snadiopt-veryclean@\texttt{snadiopt-veryclean}|see{\v{makefile} targets}}
\index{prob-clean@\texttt{prob-clean}|see{\v{makefile} targets}}
\index{phony@\texttt{phony}|see{\v{makefile} targets}}


  \bibliographystyle{siam} 
\addcontentsline{toc}{section}{References}
  \bibliography{references}

   \clearpage
   \addcontentsline{toc}{section}{Index}

   \markboth{Index}{Index}
   \begin{small}
   \printindex
   \end{small}
  \markboth{Users Guide for {\textsf SnadiOpt}}{}

\end{document}